\documentclass[eqsecnum,showpacs,amsmath,amssymb,superscriptaddress,nofootinbib,preprintnumbers]{revtex4-2}

\usepackage{graphicx}% Include figure files
\usepackage{dcolumn}% Align table columns on decimal point
\usepackage{bm}% bold math
\usepackage{color}

\newcommand{\bea}{\begin{eqnarray}}
\newcommand{\eea}{\end{eqnarray}}

\newcommand{\nn}{\nonumber\\}

\newcommand{\hide}[1]{}

\begin{document}

\preprint{NITEP 260, NU-QG-11}

\title{Bianchi-I cosmology with scale dependent $G$ and $\Lambda$ in asymptotically safe gravity}

\author{Chiang-Mei Chen} \email{cmchen@phy.ncu.edu.tw}
\affiliation{Department of Physics, National Central University, Zhongli, Taoyuan 320317, Taiwan}
\affiliation{Center for High Energy and High Field Physics (CHiP), National Central University, Zhongli,
 Taoyuan 320317, Taiwan}
\affiliation{Asia Pacific Center for Theoretical Physics (APCTP), Pohang 37673, Korea}

\author{Akihiro Ishibashi} \email{ishibashi.akihiro.r7@f.mail.nagoya-u.ac.jp}
\affiliation{Department of Physics and Kobayashi-Maskawa Institute, Nagoya University, Nagoya 464-8602, Japan}

\author{Rituparna Mandal} \email{drimit.ritu@gmail.com}
\thanks{Corresponding author}
\affiliation{Department of Physics, National Central University, Zhongli, Taoyuan 320317, Taiwan}

\author{Nobuyoshi Ohta} \email{ohtan.gm@gmail.com}
\affiliation{Nambu Yoichiro Institute of Theoretical and Experimental Physics (NITEP),
Osaka Metropolitan University, Osaka 558-5585, Japan}
\affiliation{Asia Pacific Center for Theoretical Physics (APCTP), Pohang 37673, Korea}

\date{\today}

\begin{abstract}
We study anisotropic Bianchi-I cosmology, incorporating quantum gravitational corrections into
the Einstein equation through the scale-dependent Newton coupling and cosmological term, as determined
by the flow equation of the effective action for gravity.
For the classical cosmological constant $\Lambda_0=0$, we derive the quantum mechanically corrected,
or quantum-improved power-series solution for a general
equation-of-state parameter $w$ in the range $-1<w\leq 1$ in the form of expansions in both inverse cosmic time
and the anisotropy parameter.
We give a general criterion, valid for any $\Lambda_0$, if the solution becomes isotropic in the late time, which
indicates that the universe becomes isotropic for most cases of $-1<w<1$ except $w=1$. By numerical analysis,
we show that quantum corrections lead to earlier isotropization compared
to the classical case starting from an initially highly anisotropic state.
In contrast, for $\Lambda_0 >0$, we obtain the inverse power-series solution in the exponential of the cosmic time.
We find that the universe always becomes isotropic in the late time, in accordance with the cosmic no hair
theorem, and the quantum corrections make the isotropization faster.
We also briefly summarize the Kasner solution and its generalization with quantum corrections.

\end{abstract}

%\pacs{04.62.+v, 04.70.Dy, 04.50.Gh, 04.40.Nr}

\maketitle

%%%%%%%%%%%%%%%%%%%%%%%%%%%%%%%%%%%%%%%%%%%%%%%%%%%%%%%%%%%%%%%%%%%%%%
\section{Introduction}
\label{introduction}
%%%%%%%%%%%%%%%%%%%%%%%%%%%%%%%%%%%%%%%%%%%%%%%%%%%%%%%%%%%%%%%%%%%%%%

%\section{Cosmological Model in Asymptotically Safe Gragity}

Cosmology provides a natural setting to explore possible signatures of quantum gravity, since it probes
the universe from its earliest stages through its late-time evolution. Among the various approaches,
asymptotic safety (AS), originally proposed in late 1970s~\cite{Weinber:1978, Weinberg}, attracts considerable
attention as a consistent non-perturbative framework.
Traditional perturbative quantization of gravity fails because the theory
is not renormalizable, while the effective field theory description of gravity, though useful at low energies,
ceases to be predictive near the Planck scale due to the appearance of infinitely many undetermined parameters.
The AS scenario offers a resolution by requiring that the quantum theory of
four-dimensional gravity lie within the ultraviolet (UV) critical surface of renormalization group (RG) fixed
points~\cite{Reuter:1996cp, Souma:1999at}. The non-trivial fixed points control the high-energy
regime, ensuring that dimensionless couplings remain finite.

The study of AS is quite often formulated by using the effective average action, $\Gamma_{k}[g_{\mu \nu}]$,
which encodes the physics of quantum fluctuations above the certain nonzero momentum cutoff scale $k$~\cite{Percacci:2017,
Eichhorn:2018yfc, Reuter:2019book}.
The evolution of $\Gamma_{k}$ is governed by
the functional renormalization group (FRG) equation, also referred to as flow equation~\cite{Wetterich:1992yh}:
\begin{equation}
k \frac{d\Gamma_k}{dk} = \frac12 {\rm Tr} \left\{ \left[ \Gamma_k^{(2)} + R_k \right]^{-1} k \frac{dR_k}{dk} \right\}
\label{frge}
\end{equation}
where $\Gamma_k^{(2)}$ denotes the second functional derivative of $\Gamma_{k}$, and $R_k$ is an infrared regulator.
See also~\cite{Morris:1993qb}.
The regulator suppresses fluctuations with $p^{2} \lesssim k^2$, decays rapidly for $p^{2} \gtrsim k^2$,
and vanishes as $k \to 0$, ensuring a smooth interpolation between microscopic and macroscopic physics.

A non-perturbative solution of the flow equation is obtained by employing ``truncation'', in which
the infinite-dimensional space of all possible actions is projected onto a finite-dimensional subspace
relevant to the problem under study. In the so-called Einstein-Hilbert truncation,
the RG flow of general relativity is projected onto a two-dimensional subspace:
\begin{equation}
\Gamma_k = \frac{1}{16 \pi G(k)} \int d^4 x \sqrt{-g} \left[ R(g_{\mu \nu}) - 2 \Lambda(k) \right],
\label{EHtrunc}
\end{equation}
where $R(g_{\mu \nu})$ is the Ricci scalar of the Riemannian metric $g_{\mu\nu}$, $g = \det(g_{\mu\nu})$,
$G(k)$ is the scale-dependent Newton coupling, and $\Lambda(k)$ is the running cosmological term.
Their flow in $k$ is determined by the FRG equations~\cite{Reuter:2001ag, Bonanno:2001xi, Codello:2008vh}.
This truncation to the Einstein and cosmological terms may be justified within the essential RG group
approach which removes inessential couplings by the field redefinition~\cite{Baldazzi:2021orb, Ohta:2025xxo}.

The FRG Eq.~\eqref{frge} entails that the effective average action interpolates between the fundamental action in the UV regime, and the quantum effective action as $k \to 0$. This interpolation makes the coupling `constants' scale dependent, and their evolution is governed by the FRG equation~\cite{Wetterich:1992yh,Percacci:2017,Eichhorn:2018yfc,Reuter:2019book}.
For applications to late-time cosmology, corresponding to the infrared limit $k \ll 1$,
we have shown in~\cite{Chen:2024ebb} that the scale-dependent couplings can be expanded in powers of $k^2$ as
\begin{eqnarray}
\label{eq_GLam}
G(k) &=& G_0 \left[ 1 - \omega G_0 k^2 + \omega_1 G_0^2 k^4 + \mathcal{O}\left( G_0^3 k^6 \right) \right],
\nonumber\\
\Lambda(k) &=& \Lambda_0 \left[ 1 - \mu G_0 k^2 + \mu_1 G_0^2 k^4 + \mathcal{O}\left( G_0^3 k^6 \right) \right]
 + G_0 k^4 \left[ \nu + \nu_1 G_0 k^2 + \mathcal{O}(G_0^2 k^4) \right],
\end{eqnarray}
where $G_0$ and $\Lambda_0$ are the classical Newton and cosmological constants, respectively,
and the dimensionless parameters $\omega, \omega_1, \mu, \mu_1, \nu$ and $\nu_1$ are given, for $\Lambda_0 = 0$, by
\begin{equation}
\omega = \frac{11}{6 \pi}, \quad
\omega_1 = \frac{217}{72 \pi^2}, \quad
\nu = \frac1{8 \pi}, \quad
\nu_1 = \frac{7}{54 \pi^2},
\label{data1}
\end{equation}
and for $\Lambda_0 \ne 0$, they are
\begin{eqnarray}
\omega = \mu = \frac{7}{6 \pi}
, \quad \omega_1 = \frac{49}{36 \pi^2}  - \frac{5}{24 \pi G_0 \Lambda_0},
\quad \mu_1 = \frac{49}{36 \pi^2}, \quad
\nu = - \frac{17}{24 \pi}, \quad
\nu_1=\frac{119}{144\pi^2}.
\label{data2}
\end{eqnarray}
These values were obtained by using the optimized cutoff $R_k(\Delta) \sim (k^2 - \Delta)\Theta(k^2 - \Delta)$ with suitable differential operator
$\Delta$~\cite{Codello:2008vh,Litim:2001}. It has also been shown that these may be slightly different
if we use other cutoffs, but the other choices give qualitatively similar results~\cite{Chen:2024ebb}.
So these values will be used when necessary.
Throughout this paper, we use natural units in which $\hbar = c = 1$ for convenience and the dimension of
every physical quantity may be expressed in terms of mass. For example, the dimensions of the Newton constant~$G_0$
and the cutoff scale $k$ are (mass)$^{-2}$ and (mass), respectively. In addition, only in our numerical solutions,
we choose $G_0 = 1$, corresponding to Planck unit, for simplicity.
The dimensionless parameters in \eqref{data1} and \eqref{data2} represent quantum corrections, which vanish
in the classical limit.
The quantum corrections to the cosmological term start at order $k^4$ when $\Lambda_0 = 0$, whereas
for $\Lambda_0 \neq 0$ the corrections begin at order $k^2$. The second equation in~\eqref{data2} shows that we cannot take the $\Lambda_0 \to 0$ limit smoothly since
it is singular. The phase portrait of the Newton and cosmological couplings reveals three distinct trajectories
emerging from the RG fixed point
in the UV regime: one directed toward negative cosmological constant values, another toward positive values,
and a third leading toward the origin corresponding to
$\Lambda_0 = 0$~\cite{Reuter:2001ag, ReuterWeyer2004,Codello:2008vh,Kawai:2024rau}.
The two cases therefore represent disconnected branches, and cosmological solutions with the quantum
corrections (hereafter referred to as quantum-improved solutions) must be analyzed separately for $\Lambda_0 = 0$
and $\Lambda_0 \neq 0$. Since our interest lies in an expanding universe, we restrict our attention to
$\Lambda_0 \geq 0$.

In general relativity, the Bianchi-I (BI) universe holds particular cosmological importance, as it represents
the simplest anisotropic generalization of the flat Friedmann-Lema\^{\i}tre-Robertson-Walker (FLRW) metric, which
accurately describes the present universe. However, the adequacy of the isotropic model at late times
does not guarantee its applicability to the early stages of cosmic evolution. In fact, there are theoretical
arguments suggesting that an anisotropic phase may naturally evolve into an isotropic one~\cite{Misner:1967uu}.
This motivates us to study the BI universe and its evolution toward isotropy when quantum corrections
are incorporated.

Unlike the FLRW case, where the scale factors are identical in all three spatial directions, the BI universe
is characterized by three independent scale factors. Despite its apparent simplicity, obtaining analytical
solutions of the BI dynamics, even in the classical case for a perfect fluid obeying the simplest equation
of state $p = w \rho$ with $-1<w\leq 1$ , remains challenging.
For the BI model, two well-known solutions exist: the Kasner solution for the vacuum case~\cite{Kasner:1921zz},
and the Heckmann-Sch\"ucking solution for dust~\cite{Heckmann:1959}. The latter has been further
generalized to include dust, stiff matter, and a cosmological term~\cite{Khalatnikov:2003ph, Kamenshchik:2009dt}.
The BI universe also exhibits several notable features. For matter with $p = w \rho$ and $w < 1$, the initial
anisotropy decays rapidly, and the model quickly approaches isotropy, effectively reducing to an FLRW
universe~\cite{Jacobs:1968}. Another important property is that, close to the singularity, the BI universe
asymptotically behaves like a vacuum Kasner solution, even in the presence of matter~\cite{Belinskii:1970}.
These characteristics have motivated a wide body of literature exploring BI cosmologies from different
perspectives~\cite{Hu:1978zd, Cho:1995hz, Chen:2000gaa, Saha:2001ig, Russell:2013oda}.
In Ref.~\cite{Parnovsky:2023}, a more general study of the BI homogeneous cosmological model with
matter components, including dark matter and dark energy characterized by an arbitrary barotropic equation of state,
was presented, and a possible classification of future singularities was discussed.
Also BI cosmological models with time-dependent $G$ and $\Lambda$ have been previously studied
in the presence of a perfect fluid~\cite{Beesham:1994ni, Kalligas:1995qh, Singh:2007pt, Pradhan:2013jg}
by just assuming that the $G$ and $\Lambda$ are functions of cosmic time.
Within the AS approach, the BI universe has also been examined for the $\Lambda_0 = 0$ branch under
the standard continuity equation framework~\cite{Mandal:2019xlg}.

Quantum corrections can be incorporated into cosmology at different levels, either directly at the level of
the field equations or through modifications to the gravitational action~\cite{Platania_2020}.
In recent years, several studies have proposed scale-dependent formulations of the gravitational action
that preserve diffeomorphism invariance at every scale $k$. Among these approaches is the Brans–Dicke
Lagrangian approach~\cite{Reuter:2003ca}. However, the resulting scale-dependent field equations impose
strong constraints on possible cutoff identifications, since both $G$ and $\Lambda$ generally carry energy
and momentum. In this work, we incorporate quantum effects into the Einstein equations by replacing the classical
Newton coupling and cosmological constant with their scale-dependent counterparts, as prescribed by the FRG framework.

An essential step in this procedure is the identification of the RG scale with a physical quantity of
the system. As discussed in our earlier work~\cite{Chen:2024ebb}, the Hubble parameter serves as a suitable
choice for the case $\Lambda_0 = 0$, but fails for $\Lambda_0 \neq 0$. Accordingly, we have adopted the
``improved'' classical Hubble parameter, expressed in terms of the quantum-improved scale factor,
in the $\Lambda_0 = 0$ case, while a different treatment is required when $\Lambda_0 > 0$.
Another ingredient is the modified continuity equation, derived by imposing covariant conservation
of the full left-hand side of the improved Einstein equations, distinct from the ordinary
continuity equation where the energy-momentum tensor of the fluid is conserved
separately~\cite{Bonanno:2001xi, Bonanno:2001hi, Mandal:2019xlg}. This improved approach is now widely employed in
quantum cosmology~\cite{Chen:2024ebb, Bonanno:2007wg, Reuter:2003ca, Reuter:2004nv, Hindmarsh:2011hx, Mandal:2020umo}.

For $\Lambda_0=0$, we compute a power-series solution in inverse cosmic time for a perfect fluid with
the simplest form of the equation of state $p=w \rho$ with constant $w$.
Most importantly we give the general criterion, valid for any $\Lambda_0$, if the universe becomes
isotropic in the late time, and show that it becomes isotropic for $-1<w<1$ but not for $w = 1$.
We also investigate the role of quantum corrections in the isotropization process starting from a highly
anisotropic initial state by numerically integrating the evolution equations. Our results indicate that
the quantum-improved volume expands more rapidly than in the classical case, leading to earlier isotropization.
For $\Lambda_0 > 0$, we give the inverse power-series solution in the exponential of the cosmic time
and also study the solution numerically.
We find that the isotropy is always achieved in the late time, in accordance with the cosmic no hair theorem,
ensuring that the universe asymptotically approaches a de Sitter state (for a discussion of the cosmic no hair theorem, see, e.g., \cite{Barrow:1987ia,Kitada:1991ih,Kitada:1992uh} and references therein), and the quantum corrections make the isotropization faster.

This paper is organized as follows. In Sec. II, we present the improved Einstein equations for the BI universe
and outline the method used to determine the expansion rate, the volume element, and the scale factors
for both the classical and quantum-improved cases within the AS framework.
Sec.~\ref{sec:III} focuses on the BI cosmology with $\Lambda_0 = 0$. In Subsec.~\ref{BIGS}, we discuss
some general classical and quantum solutions, and derive the classical and quantum power-series solutions
in Subsec.~\ref{power2}. Subsec.~\ref{Num0} discusses the identification of the RG scale and the choice of
initial conditions for the BI universe. We also provide numerical solutions for the volume element to
compare the classical and quantum-improved cases. In Sec.~\ref{sec:IsoBI}, we give the general criterion,
valid for any value of $\Lambda_0$, if the isotropization is achieved in the late time.
By this criterion, we show that the BI universe becomes isotropic for $-1<w<1$, but not for $w=1$ with $\Lambda_0=0$.
With $\Lambda_0>0$, we show that the BI universe always becomes isotropic in the late time.
In Sec.~\ref{BIpL}, we extend the analysis of BI universe to the $\Lambda_0 > 0$ case,
where we summarize some of the classical exact solutions and the power-series solutions, and present numerical
results for the volume element. These results indicate that the quantum corrections accelerate the isotropization.
Finally, in Sec.~\ref{summary}, we conclude with a summary of our main findings.
Appendix~\ref{kasner_app} describes the Kasner type of solutions for classical case and their quantum extensions.

%%%%%%%%%%%%%%%%%%%%%%%%%%%%%%%%%%%%%%%%%%%%%%%%%%%%%%%%%%%%%%%%%%%%%%
\section{BI Cosmology}
%%%%%%%%%%%%%%%%%%%%%%%%%%%%%%%%%%%%%%%%%%%%%%%%%%%%%%%%%%%%%%%%%%%%%%

\subsection{General discussions}
%%%%%%%%%%%%%%%%%%%%%%%%%%%%%%%%%%%%%%%%%%%%%%%%%%%%%%%%%%%%%%%%%%%%%%
In this paper, the FRG improvement is applied at the `equation level' by replacing the ordinary
Newton coupling and cosmological constant with the scale-dependent running coupling
parameters $G(k)$ and $\Lambda(k)$ in the Einstein equation. We take the viewpoint that these running
coupling parameters capture quantum effects at the leading order. The quantum improved
Einstein equation can thus be written as
\begin{equation}
R_{\mu\nu} - \frac{1}{2} R g_{\mu\nu} = - \Lambda(k) g_{\mu\nu} + 8 \pi G(k) T_{\mu\nu}.
\end{equation}
We shall identify the RG scale $k$ with a physical scale of the system, such as the Hubble parameter
or the scale factor, both of which depend on cosmic time $t$, and so do $G$ and $\Lambda$.

We consider the homogeneous but anisotropic BI cosmological model with the line element
\begin{equation}
ds^2 = - dt^2 + a_1^2(t) dx^2 + a_2^2(t) dy^2 + a_3^2(t) dz^2,
\label{Bianchi_I}
\end{equation}
where $a_1$, $a_2$ and $a_3$ represent the scale factor in three different directions which are functions of
cosmic time $t$. For this metric, the Ricci tensors and scalar curvature are given by
\begin{equation}
R_{00} = - 3 \dot H - \sum_{i=1}^3 H_i^2, \qquad R_{ii} = a_i^2 \left( \dot H_i + 3 H H_i \right), \qquad
R = 6 \dot H + 9 H^2 + \sum_{i=1}^3 H_i^2,
\end{equation}
where $H_i$ are directional Hubble parameters and $H$ is their average of the three directional Hubble
parameters defined as
\begin{equation}
H_i = \frac{\dot a_i}{a_i}, \qquad H = \frac13 \sum_{i=1}^3 H_i.
\end{equation}
The cosmic matter is assumed to be a perfect fluid, described by the energy-momentum tensor
\begin{equation}
\label{eq_EMT}
T^\mu{}_\nu = \mathrm{diag}( - \rho, p, p, p ),
\end{equation}
where $\rho$ is the energy density, $p$ is the pressure and they are related by the equation of state
\begin{equation}
\label{eq_EOS}
p = w \rho.
\end{equation}
Throughout this paper, we consider normal matter with $-1 < w \leq 1$.

For the BI metric, %together with the energy-momentum tensor,
we write these equations in terms of the directional Hubble parameter and the average Hubble parameter
\begin{eqnarray}
3 \dot H + \sum_{i=1}^3 H_i^2 &=& \Lambda - 4 \pi G (\rho + 3 p),
\label{eq:3dotH}
\\
\dot H_i + 3 H H_i &=& \Lambda + 4 \pi G (\rho - p).
\label{eq_Ein_BI_ii}
%\label{eq:dotHi}
\end{eqnarray}
We note that one may also employ a modified continuity equation, in which only the full left-hand side of
the Einstein equation is covariantly conserved. Here, the matter source $T_{\mu\nu}$ and the left-hand side
of the Einstein equation are not conserved separately, in contrast to \cite{Bonanno:2001xi}. This leads to
\begin{equation}
D^\mu \left( - \Lambda g_{\mu\nu} + 8 \pi G T_{\mu\nu} \right) = 0,
\label{Bian_iden}
\end{equation}
as a direct consequence of the Bianchi identity, $D^\mu \left( R_{\mu\nu} - \tfrac{1}{2} R g_{\mu\nu} \right) = 0$,
in Riemannian geometry.
Using Eq.~\eqref{Bian_iden}, we obtain the generalized energy conservation equation for the BI
metric~ \cite{{Bonanno:2007wg,Hindmarsh:2011hx,Reuter:2003ca,Reuter:2004nv,Mandal:2020umo}}:
\begin{equation}
\label{eq_conservation}
\dot\rho + 3 H (\rho + p) = - \frac{8 \pi \rho \dot{G} + \dot\Lambda}{8 \pi G}.
\end{equation}
The above modified conservation equation can be reproduced from the four modified Einstein
equations~(\ref{eq:3dotH}, \ref{eq_Ein_BI_ii}), but it is useful to keep in mind the difference from and similarity
to other literature.

Summing over three spatial components of the Einstein equation~\eqref{eq_Ein_BI_ii}, we get a new equation
written in terms of averaged Hubble parameter as
\begin{equation} \label{eq_Ein_BI}
\dot H + 3 H^2 = \Lambda + 4 \pi G (\rho - p).
\end{equation}
Substituting this equation back into~\eqref{eq_Ein_BI_ii} replacing the matter part, one can cast
Eq.~\eqref{eq_Ein_BI_ii} as
\begin{equation}
\dot H_i + 3 H H_i = \dot H + 3 H^2.
\end{equation}
Now, this equation can be rewritten in a more convenient form
\begin{equation}
\frac1{V} \frac{d}{dt} (V H_i) = \frac1{V} \frac{d}{dt} (V H),
\label{eq_Hi}
\end{equation}
where the volume element $V$, the multiplication of the scale factors in different directions is defined as
\begin{equation}
\label{def:V}
V := a_1 a_2 a_3 \qquad \Rightarrow \qquad \frac{\dot V}{V} = 3 H.
\end{equation}
It is pretty straightforward to calculate the directional Hubble parameters $H_i$ from Eq.~\eqref{eq_Hi}
in terms of the average Hubble parameter and the volume element $V$
\begin{equation}
\label{formula:Hi}
H_i = H + \frac{\kappa_i}{V},
\end{equation}
which in turn, leads us to the directional scale factors $a_i$ in terms of $V$ \cite{Saha:2001ig, Mandal:2019xlg}:
\begin{equation}
a_i = a_{i0} V^{1/3} \exp\left( \kappa_i \int V^{-1} dt \right),
\label{dir_scale}
\end{equation}
where the integration constants $\kappa_i$ and $a_{i0}$ satisfy relations $\sum_{i=1}^3 \kappa_i = 0$ and
$\prod_{i=1}^3 a_{i0} = 1$.
Note that the solutions for each directional Hubble parameter are determined by the integration constants $\kappa_i$.
These constants are fixed by the initial values, which are the origin of the anisotropy.

Our next task is to derive the evolution equation for the total volume $V$.
Let us first obtain from~\eqref{formula:Hi} the sum of the squares of the directional Hubble parameters
in terms of the average Hubble parameter and the volume element:
\begin{equation}
\label{sum:Hi2}
\sum_{i=1}^3 H_i^2 = 3 H^2 + \frac{\kappa^2}{V^2},
\end{equation}
where $\kappa^2 = \sum_{i=1}^3 \kappa_i^2.$
By applying the equation of state~\eqref{eq_EOS} and substituting the above expression into the Einstein
equations~\eqref{eq:3dotH} and~\eqref{eq_Ein_BI}, we obtain the evolution equations for the volume element
and the energy density in the following form
\begin{eqnarray}
&& \frac{\ddot V}{V} + \frac{w - 1}{2} \frac{\dot V^2}{V^2} - \frac{3 (w - 1)}{4} \frac{\kappa^2}{V^2}
 = \frac{3 (w + 1)}{2} \Lambda,
\label{eq_Ein_BI_V}
\\
&& \rho = \frac1{8 \pi G} \left( \frac13 \frac{\dot V^2}{V^2} - \frac12 \frac{\kappa^2}{V^2} - \Lambda \right).
\label{eq_Ein_BI_rho}
\end{eqnarray}
These two equations reduce to isotropic FLRW cosmology for $\kappa = 0$ and $V = a^3$ \cite{Chen:2024ebb,Mandal:2020umo}.

We will solve the differential equations~\eqref{eq_Ein_BI_V} for the volume $V$.
The matter density $\rho$ is then given by~\eqref{eq_Ein_BI_rho}.

\subsection{General solutions to~\eqref{eq_Ein_BI_V}}
%%%%%%%%%%%%%%%%%%%%%%%%%%%%%%%%%%%%%%%%%%%%%%%%%%%%%%%%%%%%%%%%%%%%%%

To solve Eq.~\eqref{eq_Ein_BI_V} for $V$, we define $u := \dot V$, which implies $\ddot V = u (du/dV)$.
Substituting these into Eq.~\eqref{eq_Ein_BI_V}, we rewrite the equation as
\begin{equation}
\label{eq_x}
\frac{du^2}{dV} = \frac{(1 - w) (u^2 - 3 \kappa^2/2)}{V} + 3 (1 + w) \Lambda V,
\end{equation}
which can be solved once $\Lambda$ is identified in terms of $V$.

\subsubsection{Classical solutions with constant $\Lambda_0$}

The classical solutions for the volume element $V$ %and the energy density $\rho$
can be obtained by setting $G = G_0$ and $\Lambda = \Lambda_0$.
We can then rewrite Eq.~\eqref{eq_x} as
\begin{equation}
\label{eq_x_lambda}
\frac{d( u_\mathrm{cl}^2  - 3 \kappa^2/2- 3 \Lambda_0 V_\mathrm{cl}^2)}{dV_\mathrm{cl}}
 = \frac{(1 - w) (u_\mathrm{cl}^2 - 3 \kappa^2/2 - 3 \Lambda_0 V_\mathrm{cl}^2)}{V_\mathrm{cl}},
\end{equation}
which leads to the solution for the square of the expansion rate of the volume element:
\begin{equation}
u_\mathrm{cl}^2 = c V_\mathrm{cl}^{1 - w} + 3 \kappa^2/2 + 3 \Lambda_0 V_\mathrm{cl}^2,
\label{solc1}
\end{equation}
where $c$ is an integration constant with suitable dimension. Note that the energy density~\eqref{eq_Ein_BI_rho}
in the classical solutions, upon substituting the above expansion rate of the volume element, is given by
\begin{equation}
\rho = \frac{c V_\mathrm{cl}^{-1 - w}}{24 \pi G_0}.
\end{equation}
The energy density vanishes for $c = 0$, giving the classical Kasner universe discussed in Appendix~\ref{kasner_app},
regardless of whether the cosmological term is zero or not. Thus, at least at the classical
level, the solution for the volume element in the Kasner universe is always governed by a single integration constant
and this is so because the condition $\rho = 0$ is imposed on the solution~\cite{Brizuela:2024}.

The solution for the total volume $V$ as a function of cosmic time $t$ is implicitly given by
\begin{equation}
\label{eq_tfromVc}
t - t_0 = \int_{{V}_0}^{V} \frac{dv}{u_\mathrm{cl}(v)}
\end{equation}
where ${V}_0$ is the initial volume at time $t_0$.

\subsubsection{Quantum-improved solutions}
\label{quantum}

To study quantum corrections, we need the identification of the infrared cutoff scale $k^2$ with some
time-dependent quantity.
In general, this involves relating $k$ to physical scales that are relevant to the problem under consideration.
A widely used and intuitively appealing choice in cosmological setting is to identify the cutoff with
the Hubble parameter, which naturally provides a characteristic scale in cosmology. Several studies have examined
FLRW cosmology using different choices of the cutoff scale~\cite{Reuter:2005jcap,Bonanno:2007wg,BF_2017,
Moti:2019,Hindmarsh:2011hx,Platania_2020}. In our earlier work~\cite{Chen:2024ebb}, we argued that
the late-time behaviour of the universe can be used to constrain the cutoff identification, showing that
the Hubble parameter is a suitable choice for the $\Lambda_0=0$ trajectories, but not so appropriate for
the $\Lambda_0 >0$ case. Extending this reasoning to the BI universe, the average Hubble parameter, which
characterizes the overall expansion rate of the volume, serves as a plausible choice for this identification.
This suggests
\bea
k = 3 \xi H_\mathrm{cl} = \xi \frac{\dot V_\mathrm{cl}}{V_\mathrm{cl}},
\eea
which, upon substituting~\eqref{solc1}, gives
\bea
k^2 = \xi^2 \frac{u_\mathrm{cl}^2}{V_\mathrm{cl}^2} = \xi^2 \left( c V_\mathrm{cl}^{-1-w}
 + \frac{3}{2}\kappa^2 V_\mathrm{cl}^{-2} + 3 \Lambda_0 \right),
\eea
where $\xi$ is a dimensionless constant of order one.
However this has two undesirable properties.
First of all, this identification makes the Einstein equation~\eqref{eq_Ein_BI_V} an ``inhomogeneous''
differential equation, where the running coupling $\Lambda(k(V_{\rm cl}))$ acts as a source term.
This form of the equation is generally difficult to solve analytically. In fact it is more natural, for both
physical and mathematical considerations, to consider the ``improved'' identification with $V_\mathrm{cl}(t)$
replaced by $V(t)$.
This could be regarded as including the back reactions of quantum corrections and also makes
Eq.~\eqref{eq_Ein_BI_V} a solvable ``homogeneous'' differential equation.
Second, we expect that the quantum corrections should vanish for small $k$ or large $t$ ($V \to \infty$),
but the nonzero cosmological term $\Lambda_0$ in this identification does not vanish contrary to this expectation.
The second problem is absent for $\Lambda_0 = 0$.
So we propose our identification as
\begin{equation}
k^2 = \xi^2 \left( c V^{-1 - w} + \frac{3}{2} \kappa^2 V^{-2} \right),
\label{k_iden}
\end{equation}
for $\Lambda_0 = 0$.
For $\Lambda_0 > 0$, we discuss the identification later, but it will be in terms of the volume $V$.

Equation~\eqref{eq_Ein_BI_V} now takes the form
\begin{equation}
\frac{\ddot V}{V} + \frac{w-1}{2} \frac{\dot V^2}{V^2} - \frac{3(w-1)}{4} \frac{\kappa^2}{V^2}
= \frac{3(w+1)}{2}\Lambda_0 + \sum_{i=1} \gamma_i V^{\lambda_i},
\end{equation}
where $\gamma_i$ and $\lambda_i$ are constants determined by the $\Lambda(V)$ in~\eqref{eq_GLam}
with the identification~\eqref{k_iden}. This equation can be recast as a first-order differential equation
in the following form:
\begin{equation}
\label{eq_x_Lam_gen}
\frac{d \left( u^2 - 3 \Lambda_0 V^2 - \sum_{i=1} \frac{2 \gamma_i}{\lambda_i + w + 1} V^{\lambda_i + 2} \right)}{dV}
= \frac{(1 - w) \left( u^2 - 3 \kappa^2/2 - 3 \Lambda_0 V^2 - \sum_{i=1} \frac{2 \gamma_i}{\lambda_i + w + 1}
V^{\lambda_i + 2} \right)}{V}.
\end{equation}
Solving this equation for $\lambda_i + w + 1 \neq 0$, the square of the expansion rate of the volume element
can be expressed as
\begin{equation}
u^2 = c V^{1 - w} + \frac{3}{2} \kappa^2 + 3 \Lambda_0 V^2
+ \sum_{i=1} \frac{2 \gamma_i}{\lambda_i + w + 1} V^{\lambda_i + 2}.
\label{sol_eq_x}
\end{equation}
The summation represent the quantum corrections from the running cosmological term.
The solution for the total volume $V$ as a function of cosmic time $t$ is again implicitly given by
\begin{equation}
\label{eq_tfromV}
t - t_0 = \int_{\mathcal{V}_0}^{V} \frac{dv}{u(v)}
\end{equation}
where we have chosen a different initial volume $\mathcal{V}_0$ at time $t_0$ for the case with quantum corrections.

We shall discuss the explicit solutions for $\Lambda_0 = 0$ and $\Lambda_0 > 0$ separately.

%%%%%%%%%%%%%%%%%%%%%%%%%%%%%%%%%%%%%%%%%%%%%%%%%%%%%%%%%%%%%%%%%%%%%%
\section{BI Cosmology for $\Lambda_0 = 0$}
\label{sec:III}
%%%%%%%%%%%%%%%%%%%%%%%%%%%%%%%%%%%%%%%%%%%%%%%%%%%%%%%%%%%%%%%%%%%%%%

Let us first consider the case $\Lambda_0 = 0$.

\subsection{General classical and quantum-improved solutions}
\label{BIGS}
%%%%%%%%%%%%%%%%%%%%%%%%%%%%%%%%%%%%%%%%%%%%%%%%%%%%%%%%%%%%%%%%%%%%%%

Following the discussion in Subsec.~\ref{quantum}, we set the cutoff scale as~\eqref{k_iden} and
the quantum correction of the cosmological term is
\begin{eqnarray}
\label{quant:Lambda}
\Lambda(V) &=& G_0 \tilde\nu \left( c^2 V^{-2 w - 2} + 3 c \, \kappa^2 V^{- w - 3} + \frac94 \kappa^4 V^{-4} \right)
\nonumber\\
&+& G_0^2 \tilde\nu_1 \left( c^3 V^{-3 w - 3} + \frac92 c^2 \, \kappa^2 V^{-2 w - 4} + \frac{27}4 c \,
\kappa^4 V^{- w - 5} + \frac{27}8 \kappa^6 V^{-6} \right) + \cdots,
\end{eqnarray}
where we have used $\tilde\nu = \xi^4 \nu, \tilde\nu_1 = \xi^6 \nu_1$ for convenience.

Using this expression for $\Lambda(V)$, the quantum-improved solution for the expansion rate of the volume
can be obtained from Eq.~\eqref{sol_eq_x}, and is given by ($ -1 < w \leq 1$):
\begin{eqnarray}
u^2(V) &=& c V^{- w + 1} + \frac32 \kappa^2 - G_0 \tilde\nu \left( 3 c^2 V^{-2 w} + \frac{9 (w + 1) c \,
\kappa^2}2 V^{-w - 1} + \frac{27 (w + 1) \kappa^4}{4 (3 - w)} V^{-2} \right)
\nonumber\\
&-& G_0^2 \tilde\nu_1 \left( \frac{3 c^3}2 V^{-3 w - 1} + \frac{27 (w + 1) c^2 \,
\kappa^2}{2 (w + 3)} V^{-2 w - 2} + \frac{81 (w + 1) c \, \kappa^4}{16} V^{-w - 3}
 + \frac{81 (w + 1) \kappa^6}{8 (5 - w)} V^{-4} \right) + \cdots.~~~
\label{diff1}
\end{eqnarray}
Note that we do not have to find classical solutions separately since they can be recovered by setting
$\tilde\nu = \tilde\nu_1 = 0$ in the above expression.

In what follows, we focus on approximate analytical solutions for $V(t)$  and we examine the behavior of
the difference between the quantum-improved and classical volumes as a function of cosmic time in
two special cases: $w = 1$ and $w = 0$ with the leading quantum corrections.
\begin{enumerate}
\item
For $w = 1$, Eq.~\eqref{diff1} gives
\begin{equation}
u^2(V) = (c + 3 \kappa^2/2) \left[ 1 - 3 G_0 \tilde\nu (c + 3 \kappa^2/2) V^{-2} + \mathcal{O}\left( V^{-4} \right) \right].
\label{diff2}
\end{equation}
If we ignore the higher order terms $\mathcal{O}\left( V^{-4} \right)$, the quantum-improved solution
from~\eqref{eq_tfromV} is
\bea
V_\mathrm{qu} &\approx& \left[ (c + 3 \kappa^2/2) (t - t_0)^2 + 2 \sqrt{c + 3 \kappa^2/2}
\sqrt{\mathcal{V}_0^2 - 3 G_0 \tilde\nu (c + 3 \kappa^2/2)} (t - t_0) + \mathcal{V}_0^2 \right]^{1/2}.
\label{w_1_qu_sol}
\eea
On the other hand, in the classical solution, we may impose different initial condition; $V = V_{0}$ at $t = t_0$.
We then have the classical solution
\bea
V_\mathrm{cl} = \left[ (c + 3 \kappa^2/2) (t - t_0)^2 + 2 \sqrt{(c + 3 \kappa^2/2)} V_{0} (t - t_0)
 + V_{0}^2 \right]^{1/2}
= \sqrt{c + 3 \kappa^2/2}(t-t_0)+V_0 .
\eea
In this case, each directional scale factor $a_i$ is expressed as
\bea
 a_i=a_{0i}\left[\sqrt{c + 3 \kappa^2/2}(t-t_0)+V_0 \right]^{1/3+\kappa_i/\sqrt{c + 3 \kappa^2/2}} .
\eea
Since one of the three $\kappa_i$ must have a value with the opposite sign from the other two, this solution
asymptotically describes the anisotropically expanding universe, similar to the Kasner solution in
Appendix~\ref{kasner_app}.
Note that both of these solutions have the time translation invariance and the leading term is precisely the same
since the classical term is dominant asymptotically, as expected.
To study the asymptotic behavior of the solution, let us expand this in $1/t$:
\begin{align}
V_\mathrm{qu} &\approx \sqrt{c + 3 \kappa^2/2}\ t - \left(\sqrt{c + 3 \kappa^2/2}\, t_0
- \sqrt{\mathcal{V}_0^2 -3G_0 \tilde\nu(c+3\kappa^2/2)} \right)
+ \mathcal{O}\left( t^{-1} \right).
\nn
V_\mathrm{cl} &\approx \sqrt{c + 3 \kappa^2/2}\ t - \left(\sqrt{c + 3 \kappa^2/2}\, t_0 - V_0 \right)
 + \mathcal{O}\left( t^{-1} \right).
\end{align}
The important points are (i) the coefficient of the leading term ($\sim t$) is the same as that of
the classical solution since it does not involve any quantum parameter, but
(ii) the next leading term depends on the quantum parameter
$\tilde\nu$. Thus the difference $\Delta V := V_\mathrm{qu} - V_\mathrm{cl}$ is governed by the next leading term:
\begin{equation}
\Delta V = \sqrt{\mathcal{V}_0^2 - 3 G_0 \tilde\nu (c + 3 \kappa^2/2)} - V_0,
\end{equation}
which is a constant.
Note that $\mathcal{V}_{0}^2$ should be larger than $3 G_0 \tilde\nu (c + 3 \kappa^2/2)$.
In fact, there are additional higher order corrections in $1/V$ with negative sign
in Eq.~\eqref{diff2}, $\mathcal{V}_{0}^2$ should be slightly larger than this such that the rhs of Eq.~\eqref{diff2}
is non-negative.
We shall perform the precise numerical analysis in what follows.

\item
For $w = 0$, we have
\begin{equation}
u^2(V) = c V + (3 \kappa^2/2 - 3 c^2 G_0 \tilde\nu) + \mathcal{O}\left( V^{-1} \right).
\end{equation}
Similarly to the $w=1$ case, we get the classical one
\bea
V_\mathrm{cl} = \frac{c}4 (t - t_0)^2 + \sqrt{c V_0 + 3 \kappa^2/2} (t - t_0) + {V}_0.
\label{cls}
\eea
In this case, each directional scale factor $a_i$ can be expressed as,
\begin{equation}
\label{sol:HS}
a_i = a_{i0} (t - t_+)^{1/3 + \sqrt{2/3} \kappa_i/\kappa} \cdot (t - t_-)^{1/3 - \sqrt{2/3} \kappa_i/\kappa},
% t^{p_i} \left(V_0^2 t - 1 \right)^{-p_i +2/3} \,,
\end{equation}
with $t_\pm := t_0 - (2/c)\sqrt{cV_0+3\kappa^2/2} \pm \sqrt{6} \kappa/c$ and $a_{10}a_{20}a_{30}=c/4$.
This is called the Heckmann-Sch\"ucking solution~\cite{Heckmann:1959}.
In the limit $t \rightarrow \infty$, all directional scale factors behave in the same manner as $a_i \propto t^{2/3}$
and this solution asymptotically describes the isotropically expanding flat FLRW universe.
The solution with quantum correction is
\bea
V_\mathrm{qu} \approx \frac{c}4 (t - t_0)^2 + \sqrt{c \mathcal{V}_0 + 3 \kappa^2/2
 - 3 c^2 G_0 \tilde\nu} (t - t_0) + \mathcal{V}_0,
 \label{w_0_qu_term}
\eea

Expanding these solutions~\eqref{w_0_qu_term} and \eqref{cls}, we find
\begin{equation}
\Delta V = \left( \sqrt{c \mathcal{V}_{0} + 3 \kappa^2/2 - 3 c^2 G_0 \tilde\nu}
 - \sqrt{c V_{0} + 3 \kappa^2/2} \right) t + \mathcal{O}(1).
\label{difference2}
\end{equation}
Again the leading term is canceled out from this difference.

\end{enumerate}

It should be clear now that the leading asymptotic term is determined only by the parameters $c, w, \kappa$
in the differential equation~\eqref{diff1}, but not by the initial condition.
On the other hand, the next leading term is affected by the initial condition.

\subsection{Power series solutions for $\Lambda_0 = 0$: classical and quantum-improved}
\label{power2}
%%%%%%%%%%%%%%%%%%%%%%%%%%%%%%%%%%%%%%%%%%%%%%%%%%%%%%%%%%%%%%%%%%%%%%

It is generally difficult to find exact solutions for arbitrary values of the equation-of-state parameter
$w\, (-1 < w \leq 1)$. To proceed, we attempt to derive the classical solutions of Eq.~\eqref{eq_Ein_BI_V}
for the volume element $V$ in the form of a power series in cosmic time $t$.
We assume the general ansatz in the form
\begin{equation}
V(t) = A_0 t^s \left( 1 - \frac{A_1}{t} - \frac{A_2}{t^2} + \cdots \right),
\label{gen_power}
\end{equation}
where the coefficients $A_i$ are constants determined by the Einstein equation.
From Eq.~\eqref{eq_Ein_BI_V} with $\Lambda = 0$, the characteristic equation becomes
\begin{equation}
s (s - 1) t^{-2} + \frac{w - 1}2 s^2 t^{-2} - \frac{3 (w - 1)}{4 A_0^2} \kappa^2 t^{-2 s} = 0.
\end{equation}
This yields two possible classes of solutions for the exponent $s$:
(i) $s = 1$ and the characteristic equation requires $\kappa^2 = 2 A_0^2/3$ (Kasner) with arbitrary $w$.
(ii) $s$ is greater than unity ($s > 1$, valid for $w < 1$). In this case, the term $t^{-2s}$ is higher order
than $t^{-2}$, and the characteristic equation requires $s = 2/(w + 1)$.
The power-series solution involves a ``two-dimensional" expansion in both of $1/t$ and $\kappa^2$.

For case (i) the solution can be obtained exactly which is known as Kasner solution discussed
in Appendix~\ref{kasner_app}.

%{\blue Why \eqref{gen_power} is a power series in $t$, but here is in $(t-t_0)$?}
For the case (ii), the general classical solutions, except $w = 1/3$, can be expressed as
\begin{eqnarray}
\label{sol_cl_U}
V_\mathrm{cl} &=& A_0 (t - t_0)^{2/(w + 1)} \left[ 1 - U - \frac{w (w + 1)}{2 (5 w - 3)} U^2
- \frac{w (w + 1) (4 w^2 + 5 w - 3)}{6 (5 w - 3) (7 w - 5)} U^3 \right.
\nonumber\\
&& \left. - \frac{w (w + 1) (162 w^5 + 351 w^4 - 269 w^3 - 457 w^2 + 423 w - 90)}
{24 (5 w - 3)^2 (7 w - 5) (9 w - 7)} U^4 + \cdots \right],
\end{eqnarray}
where we have defined a new variable
\begin{equation}
U = - \frac{3 (1 + w)^2 \kappa^2}{8 (3 w - 1) A_0^2} (t - t_0)^{2 (w - 1)/(w + 1)}.
\end{equation}
There are three special cases:
\begin{enumerate}
\item [(a)]
For $w = 1/3$, $U$ diverges, and this case is a particular isometric solution
of $\kappa^2 = 0$ and $V_\mathrm{cl} = A_0 (t - t_0)^{3/2}$,

\item [(b)]
For $w = 1$, $U$ becomes constant, and the solution is exact as $V_\mathrm{cl} = A_0 (t - t_0)$,

\item[(c)]
For $w = 0$, all coefficients of $U^j$ for $j \ge 2$ vanish,
leading again to an exact solution, given by $V_\mathrm{cl} = A_0 (t - t_0)^2 - {3 \kappa^2}/{8 A_0}$.

\end{enumerate}

As the governing equation~\eqref{eq_Ein_BI_V} is a second-order differential equation, it admits two integration
constants, $A_0$ and $A_1$. All other constants in Eq.~\eqref{gen_power} can be expressed in terms of these two.
In the classical solution above, we have redefined $A_1$ in terms of $t_0$ as $A_1 = - 2 t_0 /(w + 1)$, so that
this constant can be interpreted on the same footing as the cosmic time $t$. From the above solution,
we could expand the above solution in the power series of both $1/t$ and $\kappa^2$ at late time.
Below, we present the quantum-improved solution for the volume element using the cutoff identification
in Eq.~\eqref{k_iden}, starting from~\eqref{eq_Ein_BI_V} and expressing it as a power series in both parameters.

The general power series solution of quantum-improved $V(t)$, for $w \ne 1$ and $w \ne 1/3$, is
\begin{eqnarray}
\label{eq_V_cl_series}
V(t) &=& A_0 t^{\frac2{w + 1}} \left\{ 1 - \frac{2}{w + 1} \frac{t_0}{t}
- \frac{w - 1}{(w + 1)^2} \frac{t_0^2}{t^2} - \frac{2 w (w - 1)}{3 (w + 1)^3} \frac{t_0^3}{t^3}
- \frac{w (w - 1) (3 w + 1)}{6 (w + 1)^4} \frac{t_0^4}{t^4}  \right. \nn
&+& \frac{3 (w + 1) \tilde\nu G_0 c^2}{4 A_0^{2 w + 2} t_0^2} \frac{t_0^2}{t^2}
+ \frac{3 w \tilde\nu G_0 c^2}{2 A_0^{2 w + 2} t_0^2} \frac{t_0^3}{t^3}
+ \left. \left[  \frac{3 w (3 w + 1)\tilde\nu G_0 c^2}{4 (w + 1) A_0^{2 w + 2} t_0^2}
\!+\! \frac{(w \!+\! 1) \tilde\nu_1 G_0^2 c^3}{8 A_0^{3 w + 3} t_0^4}
\!-\! \frac{9 w (w \!+\! 1)^2 \tilde\nu^2 G_0^2 c^4}{32 A_0^{4 w + 4} t_0^4} \right] \frac{t_0^4}{t^4}
+ \mathcal{O}(t^{-5}) \right\},
\nn
&+& \frac{3 (w + 1)^2 \kappa^2}{8 (3 w - 1) A_0} t^{\frac{2 w}{w + 1}} \left\{ 1 - \frac{2 w}{w + 1}
\frac{t_0}{t} + \left[ \frac{w (w - 1)}{(w + 1)^2}
\!-\! \frac{3 (w \!+\! 1) \tilde\nu G_0 c \left[ 4 (3 w \!-\! 1) A_0^{w + 1}
\!-\! w (5 w \!+\! 1) c \right]}{4 (w - 3) A_0^{2 w + 2} t_0^2} \right] \frac{t_0^2}{t^2} + \mathcal{O}(t^{-3}) \right\}, \nn
 &+& \mathcal{O}\left( \frac{\kappa^4}{A_{0}^{4}} t^{\frac{2(2w-1)}{(w+1)}} \right)
\end{eqnarray}
Note that, by setting the quantum parameters to zero, we recover the late-time classical solution~\eqref{sol_cl_U}.
Indeed, the first line of Eq.~\eqref{eq_V_cl_series} can be identified as the leading contribution in
Eq.~\eqref{sol_cl_U}, since it represents the Taylor expansion of $(1 - x)^{2/(w+1)}$, $x = t_0/t$:
\begin{equation*}
(1 - x)^{\frac2{w + 1}} = 1 - \frac{2}{w + 1} x - \frac{w - 1}{(w + 1)^2} x^2
- \frac{2 w (w - 1)}{3 (w + 1)^3} x^3 - \frac{w (w - 1) (3 w + 1)}{6 (w + 1)^4} x^4
- \frac{w (w - 1) (2 w + 1) (3 w + 1)}{15 (w + 1)^5} x^5 + \cdots.
\end{equation*}
Similarly, the first three terms in the last line of Eq.~\eqref{eq_V_cl_series} (the part proportional to $\kappa^2$)
correspond to the Taylor expansion of $(1 - x)^{2 w/ (w + 1)}$:
\begin{equation*}
(1 - x)^{\frac{2 w} {w + 1}} = 1 - \frac{2 w}{w + 1} x + \frac{w (w - 1)}{(w + 1)^2} x^2
+ \frac{2 w (w - 1)}{3 (w + 1)^3} x^3 + \frac{w (w - 1) (w + 3)}{6 (w + 1)^4} x^4
+ \frac{w (w - 1) (w + 2) (w + 3)}{15 (w + 1)^5} x^5 + \cdots.
\end{equation*}
From the above expressions, we can also infer the higher-order terms of the classical solution by expanding
the series. However, this is not as straightforward in the quantum case. It is true that one can write
the solution in the $(t - t_0)$ form with some exponent, as in Eqs.~(\ref{w_1_qu_sol},~\ref{w_0_qu_term}),
incorporating only the leading corrections to the expansion rate, since the Einstein equation with
our choice of cutoff preserves time-translation symmetry. Nevertheless, it is difficult to express
in a time-translation-invariant form when all terms are included in Eq.~\eqref{diff1}.

For $w = 1$, the $\kappa^2$ part (the third line of solution~\eqref{eq_V_cl_series}) should be replaced by
\begin{equation}
\kappa^2 \left\{ U_0 + \frac{9 \tilde\nu G_0 (3 \kappa^2 + 4 c)}{8 A_0^3} \frac1{t}
+ \frac{3 \tilde\nu G_0}{8 A_0^4} \left[ 3 A_0 t_0 (3 \kappa^2 + 4 c)
- U_0 (3 \kappa^2 + 2 c)^2 \right] \frac1{t^2} + \mathcal{O}(t^{-3}) \right\}.
\end{equation}

The case  $w = 1/3$ is more subtle. The implicit solution from~\eqref{diff1}, contains a logarithmic term,
$$ \frac23 c^{3/2} (t - t_0) = \sqrt{c} V^{1/3} \sqrt{c V^{2/3} + 3 \kappa^2/2}
- \frac{3\kappa^2}2 \ln\left( \sqrt{c} V^{1/3} + \sqrt{c V^{2/3} + 3 \kappa^2/2} \right). $$
Then the power series expansion involves also logarithmic terms, but this is a special class of solution
which is of little interest for the general properties we are studying. So we do not expose it explicitly here.
Nevertheless, we numerically solve this case as well and observe that the difference between the quantum-improved and classical solutions behaves in the same way as for other values of $w$.
This occurs for the same reason discussed below Eqs.~\eqref{difference2} and~\eqref{leadinga}.

Two comments are in order:
\begin{itemize}

\item
The solution is composed of two ``dimensional'' series, in powers of $1/t$ and $\kappa^2$. The isotropic
part begins at order $2/(w + 1)$, the $\kappa^2$-part at order $2 w/(w + 1)$,
$\kappa^4$-part at order $2 (2 w - 1)/(w + 1)$, and so on. These two leading orders equal to each other
for $w = 1$ and the $\kappa^2$ part appears higher order when $w < 1$.

\item
The $\kappa^2$-series can be included in the first part when the difference between the leading orders
is an integer, i.e., when $2 (w - 1)/(w + 1) \in Z \to w = 1, 1/3, 0, -1/5, -1/3, -3/7, \cdots$.

\end{itemize}

\subsection{Numerical solutions for any $w$}
\label{Num0}

The above series expansions of the solutions are useful to find the qualitative difference between the classical
and quantum-imroved solutions but cannot tell it quantitatively since they are valid only asymptotically
for large $t$.
To see how they differ quantitatively for general $w$, it is convenient to numerically integrate the differential
equation~\eqref{diff1} by writing it as
\begin{equation}
\dot V(t) = u(V(t)).
\end{equation}
and imposing suitable initial conditions. Here the right-hand side should be understood as the square root of
the right-hand side of Eq.~\eqref{diff1}.

\subsubsection{Choice of the initial condition}
\label{choice}
%%%%%%%%%%%%%%%%%%%%%%%%%%%%%%%%%%%%%%%%%%%%%%%%%%%%%%%%%%%%%%%%%%%%%%

\begin{figure}[t]
\includegraphics[scale=0.4, angle=0]{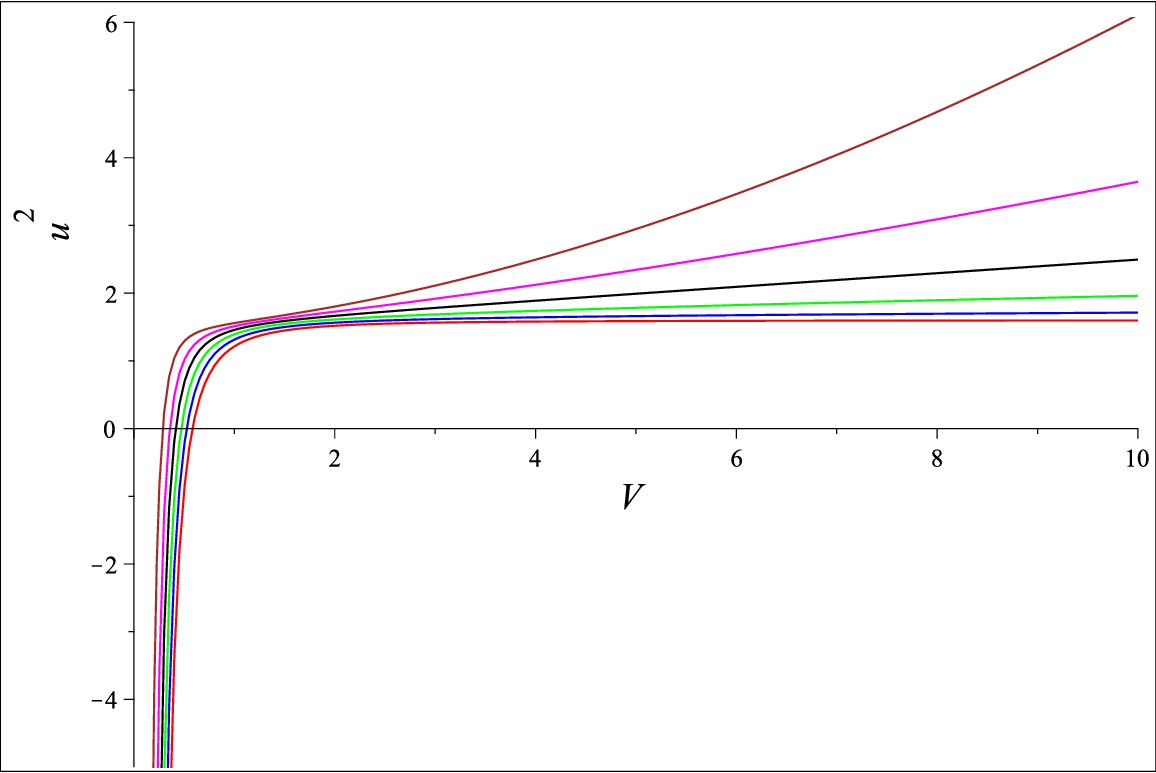}
\caption{The time derivative of volume, $u^2 = \dot V^2$ with quantum correction for $w = 1$ (red),
$w = 2/3$ (blue), $w = 1/3$ (green), $w = 0$ (black), $w = -1/3$ (magenta) and $w = -2/3$ (brown)
for the the parameter values from Eq.~\eqref{data1} and $c = 0.1, G_0 = 1, \kappa = 1$, $\xi = 1$.}
\label{f0}
\end{figure}

Now we have to specify the initial condition at some time, say $t = t_0 = 0$.
It is true that our approximation for the quantum case might not describe the evolution from the
``big bang'' at $t = 0$, but we may still give rough estimate of the volume evolution starting slightly away from it.
With this expectation, we can estimate the initial conditions from the property of the differential
equation~\eqref{diff1}.
For classical solutions ($\tilde\nu = \tilde\nu_1 = 0$), the initial value of the volume can be chosen as
$V_0 = 0$ which is the smallest possible value for the volume corresponding to the big bang.
However, for the quantum-improved case, quantum effects give negative corrections in $u^2$ as shown
in~\eqref{diff1}. Indeed we can confirm this from Fig.~\ref{f0} which shows the behavior the right-hand side of
Eq.~\eqref{diff1} for various $w$. Since $u^2$ has to be non-negative, the initial value $\mathcal{V}_0$
for quantum-improved case have to be at the point $u^2 \geq 0$.
Otherwise $u(V)$ is imaginary and unphysical.
This may be considered the reflection of quantum uncertainty principle.
The volume at that time can be estimated by the smallest value of the volume which is given
when the rhs of~\eqref{diff1} is zero. So it is natural to set the initial condition as
\begin{equation}
\left \{
\begin{array}{l}
V_{\rm cl}(0) = V_{0} = 0 \\
V_{\rm qu}(0) \sim \mathcal{V}_0
\end{array}
\right. \qquad \mathrm{at} \quad t = 0,
\label{initialc}
\end{equation}
and we determine the initial volume $\mathcal{V}_0$ such that $u(\mathcal{V}_0) \sim 0$.

\subsubsection{Evolution of the volume for $-1 < w \leq 1$}
\label{power1}
%%%%%%%%%%%%%%%%%%%%%%%%%%%%%%%%%%%%%%%%%%%%%%%%%%%%%%%%%%%%%%%%%%%%%%

The differential equation~\eqref{diff1} to be solved contains only the powers of volume though the powers
are not rational numbers in general.
Still the solutions can be given generally by power series in the cosmic time $t$.
It may be divided into the series in $t$ and $\kappa^2$.
The general power series solutions are given in~\eqref{eq_V_cl_series} and the first three essential leading terms are
\begin{equation}
V \sim A_0 t^{2/(w + 1)} - \frac{2 B}{w + 1} t^{(1 - w)/(1 + w)}
+ \frac{3 (w + 1)^2 \kappa^2}{8 (3 w - 1) A_0} t^{2 w/(w + 1)} + \cdots,
\label{asymptotic}
\end{equation}
where $A_0$ is some number determined by the parameters in the differential equation,
and $B$ is a constant determined by the initial condition.
For the values $-1 < w \le 1$, the leading terms is $t^{2/(w + 1)}$ and the subleading term is
either $t^{(1 - w)/(1 + w)}$ for $w < 1/3$ or $t^{2 w/(1 + w)}$ for $w > 1/3$.
It is actually possible to determine the coefficient $A_0$ for $-1 < w < 1$ from Eq.~\eqref{diff1}:
\bea
A_0 = \left( \frac{(w+1)^2}{4} c \right)^{1/(w+1)}.
\label{leadinga}
\eea

We again note that the parameter $A_0$ only depends on $c$ and $w$, and both the first and third terms
in the power series solutions~\eqref{asymptotic} cancel out in $\Delta V (= V_\mathrm{qu} - V_\mathrm{cl})$.
The leading term is due to the classical effect, which is present in both classical and quantum-improved volume evolutions and disappear in the difference.
The quantum effects come into the expansion in the next-to-leading terms and higher
and affect the second term in~\eqref{asymptotic} with $B$ which also depends on the initial conditions
and gives the dominant contribution to $\Delta V$.

\begin{figure}[t]
\begin{minipage}{80mm}
\includegraphics[scale=0.4, angle=0]{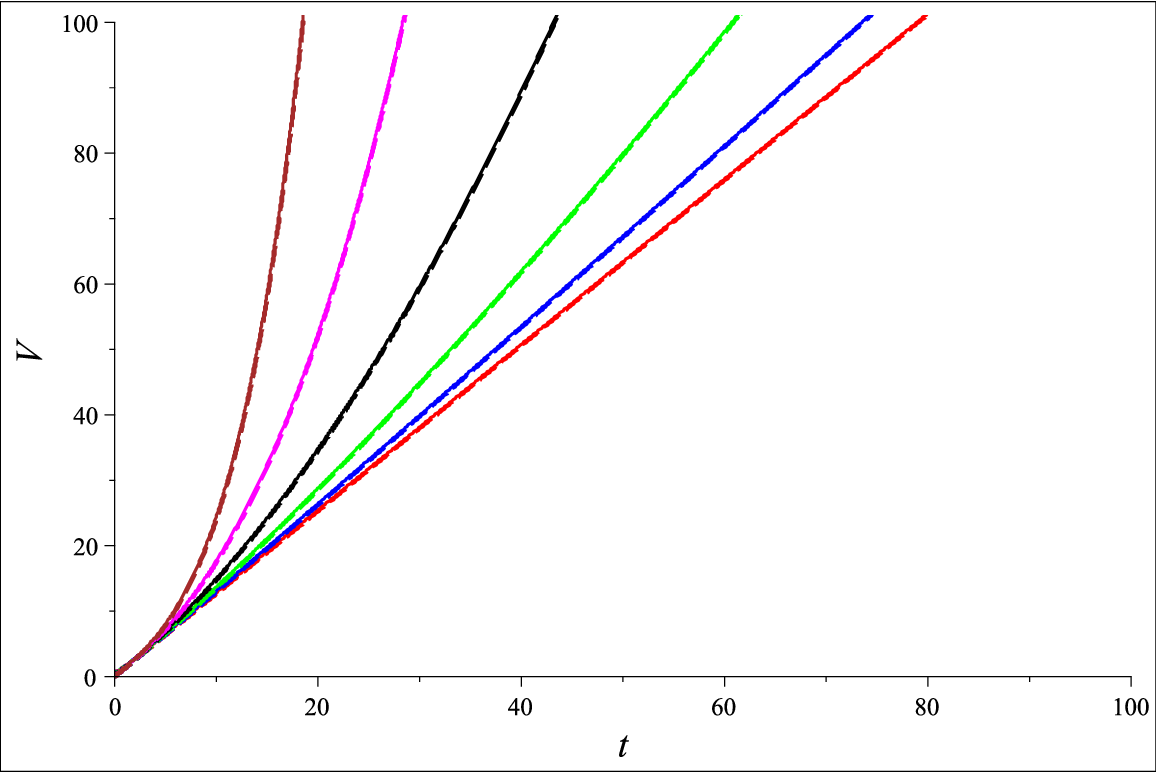}
%\vs{15}
\caption{The evolution of quantum-improved (solid line) and classical (dashed line) volumes for $w = 1$ (red),
$w = 2/3$ (blue), $w = 1/3$ (green), $w = 0$ (black), $w = -1/3$ (magenta) and $w = -2/3$ (brown)
for the chosen parameters.}
\label{f1}
\end{minipage}
\hspace{.5cm}
\begin{minipage}{80mm}
\includegraphics[scale=0.4, angle=0]{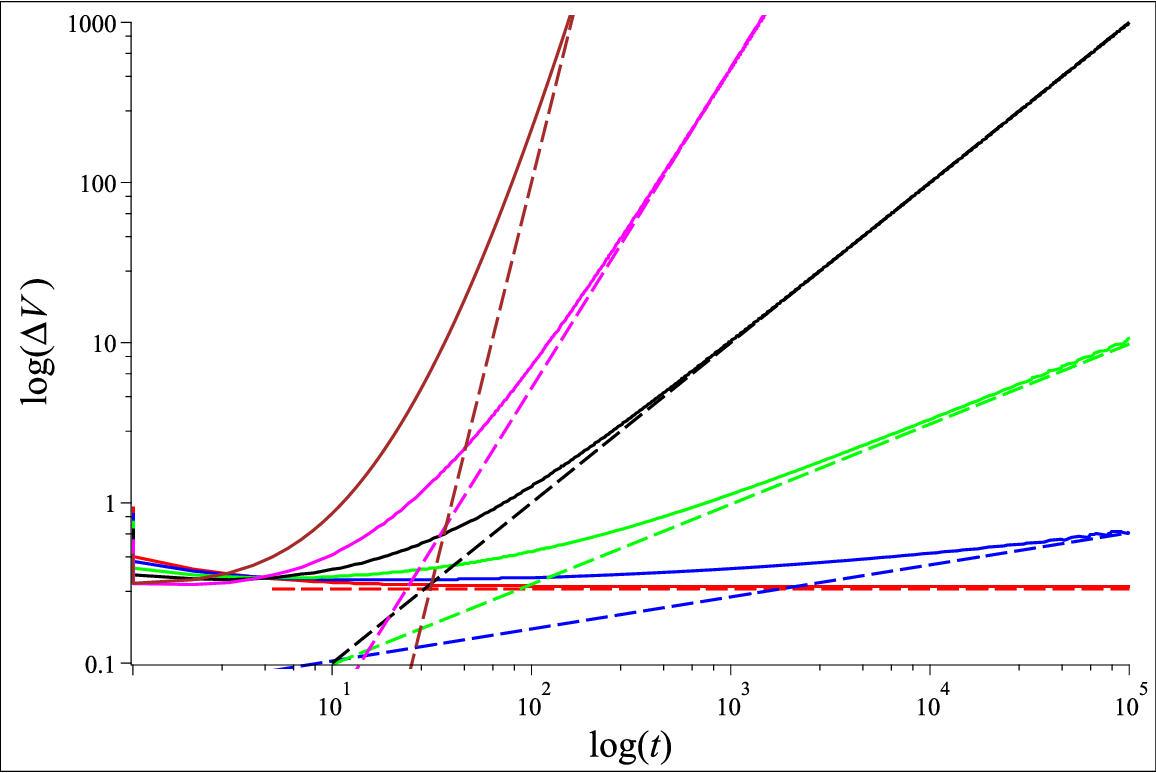}
%\vs{15}
\caption{The corresponding volume difference $\Delta V = V_\mathrm{qu} - V_\mathrm{cl}$,
in logarithmic scales,
for the chosen parameters. The dashed line is the asymptotic behaviors given in Table~\ref{t1}.
}
\label{f2}
\end{minipage}
\end{figure}

Imposing the initial condition~\eqref{initialc}, we numerically integrate Eq.~\eqref{diff1} to find
the time dependence of $V$ with the parameter values from Eq.~\eqref{data1} and
$c = 0.1, G_0 = 1, \kappa = 1$, $\xi = 1$.
We show the result in Fig.~\ref{f1}.
The initial conditions for the quantum-improved case are given in Table~\ref{t1}, together with
the leading behaviors of the total volume in $t$, which are the same for both classical and {quantum-improved} cases,
and these behaviors are in agreement with the leading terms in Eq.~\eqref{asymptotic}.
We find that the quantum-improved volume is always larger
than the classical volume throughout the later evolution. This means that the quantum effects make the expansion
faster than the classical case. The physical reason is that the quantum Newton coupling is always smaller
than the classical value, so if our universe starts expanding from some volume at early time when quantum effects
are in work, this makes the volume expanding faster than the classical case.

We also depict the difference between the {quantum-improved} and classical volumes
$\Delta V = V_\mathrm{qu} - V_\mathrm{cl}$
in Fig.~\ref{f2}. It is expected to be given by the next-to-leading term in~\eqref{asymptotic}.
Indeed we see this behavior for $w = 1, 2/3, 1/3, 0, -1/3, -2/3$, as given in the figure and Table~\ref{t1}
for the same initial conditions.

\begin{table}[tb]
\begin{center}
\begin{tabular}{|c||c|l|l|}
\hline
\quad value of $w$ \quad & \quad initial volume $\mathcal{V}_0$ \quad & \quad asymptotic behavior \quad
& \quad difference of volume $\Delta V$ \quad \\
\hline\hline
$1$ & $0.582693$ & $V \sim 1.26\, t$ & $\Delta V \sim 0.175 \quad \mbox{(constant)}$
\\ \hline
$2/3$ & $0.525390$ & $V \sim 2.9 \times 10^{-1} \, t^{6/5}$ & $\Delta V \sim 4 \times 10^{-2} \, t^{1/5}$
\\ \hline
$1/3$ & $0.470843$ & $V \sim 1.15 \times 10^{-1} \, t^{3/2}$ & $\Delta V \sim 2 \times 10^{-2} \, t^{1/2}$
\\ \hline
$0$ & $0.416325$ & $V \sim 2.6 \times 10^{-2} \, t^2$ & $\Delta V \sim 6.2 \times 10^{-3} \, t$
\\ \hline
$-1/3$ & $0.358023$ & $V \sim 1.2 \times 10^{-3} \, t^3$ & $\Delta V \sim 3.5 \times 10^{-4} \, t^2$
\\ \hline
$-2/3$ & $0.286334$ & $V \sim 2.4 \times 10^{-8} \, t^6$ & $\Delta V \sim 4 \times 10^{-9} \, t^5$
\\ \hline
\end{tabular}
\end{center}
\caption{Asymptotic behaviors of the total volume and the difference for the same initial conditions}
\label{t1}
\end{table}

Here, all equations are solved in Planck units, as the parameters $c$, $\kappa$ and the cosmic time $t$
would correspond to extremely large values in SI units.
Since $\kappa$ has the physical dimension of inverse time, its value in SI units can be estimated by dividing it
by the Planck time, yielding $\kappa = 1/(5 \times 10^{-44} ~\text{s})$. This provides a sense of the scale of
the anisotropy parameter $\kappa$. In principle, the specific choices of $c$ and $\kappa$ depend on the initial
conditions of the universe. The primary constraint on selecting these parameters is that the subleading quantum
correction terms in Eq.~\eqref{diff1} must remain smaller than the leading-order quantum terms.
While the chosen initial values might suggest that the subleading term $\kappa^6/V^4$ is larger than
the leading term $\kappa^4/V^2$, this apparent discrepancy arises only at early times when $V(t) \ll 1$.
Since our analysis focuses on the late-time behavior of the universe, where the volume element becomes
significantly larger, these parameter choices remain valid and are sufficient to capture the dominant
quantum effects in that regime.

\begin{figure}
\includegraphics[scale=0.4, angle=0]{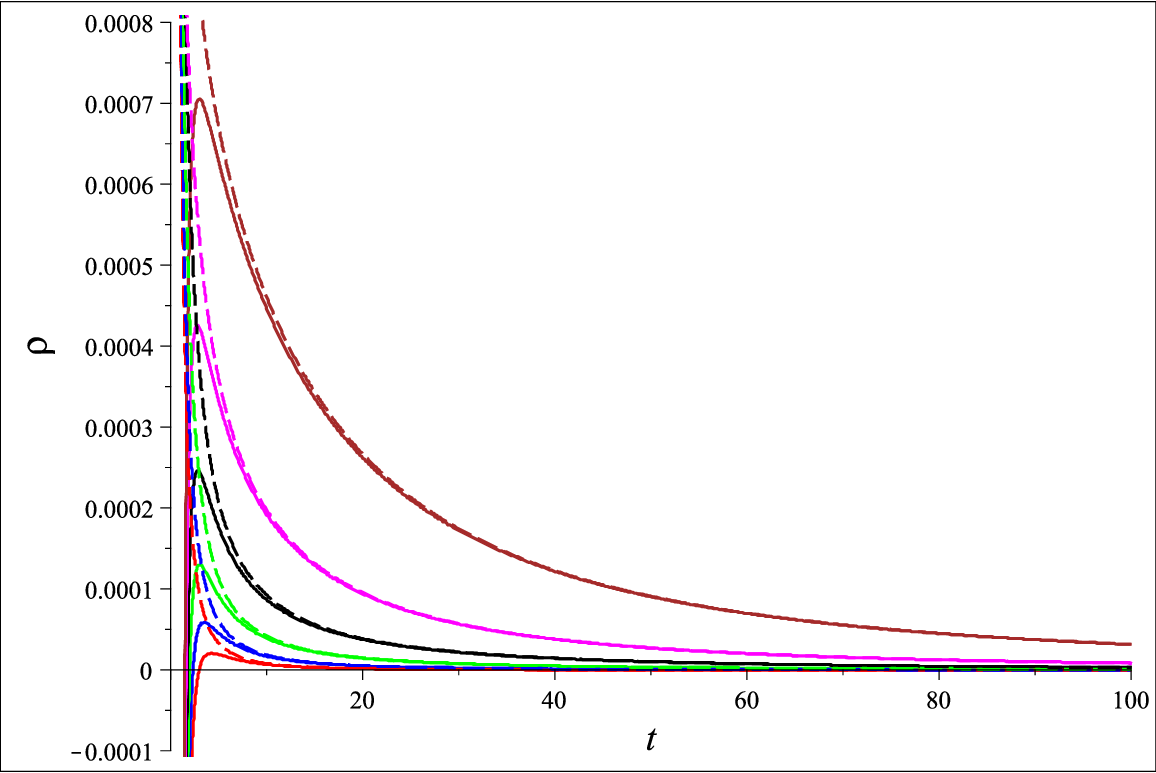}
\caption{The evolution of energy density of quantum-improved (solid) and classical (dashed)
solutions for $w = 1$ (red),
$w = 2/3$ (blue), $w = 1/3$ (green), $w = 0$ (black), $w = -1/3$ (magenta) and $w = -2/3$ (brown)
for the chosen parameters.}
\label{f3}
\end{figure}

According to the choice of initial condition~\eqref{initialc}, the energy density~\eqref{eq_Ein_BI_rho} of
classical solution diverges at $t = 0$ with initial condition of zero volume. For the quantum-improved solution,
the energy density starts from negative value due to the initial condition of vanishing time derivative of volume,
then it becomes positive and quickly reduces to classical values. The numerical results are given in Fig.~\ref{f3}.

\section{Isotropization of BI models into FLRW universe}
\label{sec:IsoBI}
%%%

The isotropic and homogeneous universe we observe today, described by the FLRW model, may have emerged as
an asymptotic state from an initially highly anisotropic universe. In the anisotropic BI model, we explore
how quantum corrections can influence the process of isotropization.

\subsection{General consideration for $\Lambda_0 = 0$}
\label{VA}

Let us first consider the case $\Lambda_0 = 0$ and study whether such an isotropization can occur.
To illustrate the nature of this problem, let us briefly look at the general
behavior of the cosmic shear and the Hubble parameter.

We assume that our BI universe is initially expanding: $H > 0$ at $t = t_0$. At this stage,
by ``expanding'' we mean that the $3$-volume $V(t) = a_1 a_2 a_3$ is increasing with cosmic time $t$,
but it does not necessarily mean that all scale factors $a_i$ are increasing with $t$. As a physically
reasonable condition, we impose (a slightly stronger version of) the dominant energy condition, which requires
$\rho > |p| \geq 0$ for our energy-momentum tensor~\eqref{eq_EMT} and~\eqref{eq_EOS}, hence $-1 < w \leq 1$,
and also the strong energy condition, i.e., $1 + 3 w > 0$.
Let $n^\mu \partial_\mu = \partial_t$ be the unit normal vector to $t = \mathrm{const.}$
hypersurface $\Sigma_t$ with the induced metric $g_{ij}$. Then, by using the formula~\eqref{formula:Hi},
the extrinsic curvature $K_{ij} := \nabla_i n_j $ of $\Sigma_t$ is expressed as
\begin{equation}
K_{ij} = H_i g_{ij} = \left( H + \dfrac{\kappa_i}{V} \right) g_{ij} \,.
\end{equation}
In our metric ansatz, each fluid element follows the timelike geodesic orthogonal to $\Sigma_t$,
and thus the shear $\sigma_{ij}$ of the fluid is given by
\begin{equation}
 \sigma_{ij} := K_{ij} - \dfrac{1}{3} K g_{ij} = \dfrac{\kappa_i}{V} g_{ij} \,,
\end{equation}
where $K := g^{ij} K_{ij}$. In particular, the shear-squared is expressed as
\begin{equation}
\label{shear2}
 \sigma_{ij} \sigma^{ij} = \dfrac{\kappa^2}{V^2} \,.
\end{equation}

By using~(\ref{eq:3dotH}) and~(\ref{sum:Hi2}), we have for $\Lambda_0 = 0$ case,
\begin{equation}
\label{eq:Raychaudhuri:0}
 \dot{H} = - H^2 -\dfrac{\kappa^2}{3V^2} %+ \dfrac{\Lambda_0}{3}
 - \dfrac{4\pi G}{3} (1 + 3 w) \rho \,.
\end{equation}
By noting that the second term in the right-hand side is the shear-squared~(\ref{shear2}),
this is identified with the Raychaudhuri equation for timelike geodesic congruence with the tangent $n^a$.
Then, combining this equation with~(\ref{eq_Ein_BI}), we obtain the constraint equation,
\begin{equation}
\label{eq:init:constraint:0}
 H^2 = % \dfrac{\Lambda_0}{3} +
 \dfrac{\kappa^2}{6 V^2} + \dfrac{8 \pi G}{3} \rho \,.
 % K^2 = 3\Lambda_0 + \dfrac{3}{2} \sigma_{ij} \sigma^{ij} +24 \pi G \rho \geq 0 \,. % 3 \Lambda_0 > 0   \,,
\end{equation}
Then, as far as initially anisotropic $\kappa^2 > 0$ or non-empty $\rho > 0$,
the constraint~(\ref{eq:init:constraint:0}) implies $H^2 > 0$ and the signature of $H$ never changes.
Therefore, if the universe is initially expanding, $H > 0$ at $t_0$, it is expanding $H > 0$ for any $t > t_0$.
Furthermore, since $\dot{V} = 3 V H > 0$, $V$ is a monotonically increasing function, and therefore
the shear squared given by~(\ref{shear2}) is a monotonically decreasing function\footnote{%%%
This can be seen for example, for the Kasner case $V = \ell_0^3 \sinh(3 t/2 \ell_0) \cosh(3 t/2 \ell_0)
\rightarrow (\ell_0 e^{t/\ell_0})^3$ as $t \rightarrow \infty$ with $\ell_0^2 = 3/\Lambda_0$.
}. %%% footnote
As a consequence of the strong energy condition, from eqs.~(\ref{eq:Raychaudhuri:0}) and~(\ref{eq:init:constraint:0}),
we have
\begin{equation}
\dot{H} < - H^2 < 0 \,. %\dot{K} \leq - \dfrac{1}{3}K^2 \leq 0 \,.
\end{equation}
%implying that $K=\sum_i H_i $ is now a monotonically decreasing function.
Integrating this inequality yields
\begin{equation}
0 < H < \dfrac{1}{t - t_0} \,, %0< {K} \leq \dfrac{3}{t} + K_0 \,,
\label{ineq:noLambda}
\end{equation}
with some integration constant $t_0$. Therefore, we find,
\begin{equation}
0 < H_1 + H_2 + H_3 < \dfrac{3}{t - t_0} \; \rightarrow \; 0 \quad (\mbox{as $t \rightarrow \infty$})\,.
\end{equation}
This implies that the following two cases can occur: (i) all $H_i$ approach zero with all $H_i$ being kept
positive, and (ii) all $H_i$ approach zero with two of $H_i$, say $H_1$ and $H_2$, having the same sign,
while the third $H_3$ the opposite sign with $\sum_i H_i > 0$. The case (i) corresponds to
the isotropization, i.e., the universe expands in all $3$ directions and eventually approaches
a flat FLRW universe. %e, as in the case of Heckmann-Sch\"ucking solution given in (\ref{sol:HS}).
As for the case (ii), suppose $H_1$ and $H_2$ are positive while $H_3$ is negative.
Then, the BI universe in case (ii) is contracting along $x^3$ direction in the asymptotic
future $t \rightarrow \infty$, though the $3$-volume $V$ itself is increasing.
It follows from~(\ref{eq:init:constraint:0}) that $8\pi G \rho \leq H^2 \rightarrow 0$ as $t \rightarrow \infty$,
hence the universe approaches a vacuum spacetime in the asymptotic future. Therefore, the case (ii)
approaches the vacuum Kasner solution discussed in Appendix~\ref{kasner_app}.

\subsection{General criterion for isotropization}

To study the isotropization in more detail, we examine how the anisotropy parameters become negligible and
how the three directional scale factors grow at the same rate as the universe evolves to the far future.

Looking at our general solution~\eqref{dir_scale}, we notice that the anisotropy is determined by the last factor.
If this factor becomes larger and larger in $t$, it is clear that the anisotropy does not vanish.
However, if this factor converges, the solution can be made isotropic.
We propose our criterion for the isotropization that the following integral is finite:\footnote{
A different criterion
$
a_i/V^{1/3} \rightarrow \text{constant} > 0$ when $t \rightarrow \infty
$
is proposed in~\cite{Bronnikov:2004nu, Saha:2006iu}, which depends on the directions.
We believe that our criterion is better because it does not depend on the direction.}
\bea
\int_{t_0}^\infty V^{-1} dt =: C_0 < \infty.
\label{criterion}
\eea
It may appear that the scale factors $a_i/V^{1/3} \rightarrow a_{i0} \, \mathrm{e}^{\kappa_i C_0}$ for
$t \rightarrow \infty$,
and hence anisotropy remains. However, we can rescale the coordinates to set these values to the same value,
and the resulting dynamics become equivalent to those of an FLRW universe.
On the other hand, if this quantity diverges as the time becomes larger, we cannot absorb the anisotropy
just by rescaling of the coordinates. We emphasize that this criterion is valid not only for $\Lambda_0 = 0$
but also $\Lambda_0 > 0$.

Now how is the isotropy realized in our present setting?
It is clear that whether the integral~\eqref{criterion} converges or not is governed by the asymptotic behavior in $t$
of the volume $V$; since we start with the finite value $V_0$ of the initial condition at $t_0$ and the total
volume is becoming large, the lower finite integration region does not give divergent contribution.

\subsubsection{$\Lambda_0 = 0$}

For $\Lambda_0 = 0$, we have seen that our solutions are given asymptotically
in terms of power of $t$. Its leading behavior is given by $t^{2/(w+1)}$. Since we are considering
$-1 < w \leqq 1$, we have
\bea
\frac{2}{w+1} \geqq 1.
\eea
So the only dangerous case with anisotropy is $w = 1$, when we have the behavior $V \sim A \, t$
(as given in~\eqref{asymptotic}), and the integral gives
$\log(t)$. In this case, our solutions are
\bea
a_i \approx a_{i0}\, t^{1/3 + \kappa_i/A},
\label{scalei}
\eea
so the inisotropy persists.
The value $w = 1$ corresponds to stiff matter, and the solution for $V(t)$ could flow to the vacuum Kasner universe
($\rho = 0$) depending on the behavior of the directional Hubble parameter $H_i$.
Since $\sum_i \kappa_i = 0$, there are cases when one of the directions shrink but others expand and/or
all three directions expand depending on $\kappa_i/A + 1/3$ is positive or not.
There is no case that all three directions shrink.
Also Eq.~\eqref{scalei} does not imply that the solution is always anisotropic.
If all the $\kappa_i$ are equal, it becomes isotropic after suitable scaling of the coordinates.

On the other hand, if $-1 < w < 1$, the exponent of the leading behavior is always larger than 1:
\bea
\frac{2}{w+1} > 1,
\eea
and the integral~\eqref{criterion} converges and our solution always asymptotically becomes isotropic.
From Fig.~\ref{f2}, we see that the quantum-improved volume is larger than the classical one,
$V(t) > V_\mathrm{cl}(t)$ at a given time, giving the rapider growth of the BI universe
results in the earlier isotropization than the classical solution due to the quantum corrections.

\subsubsection{$\Lambda_0 > 0$}

For $\Lambda_0 > 0$, we will see that our expanding solutions always have exponential type of expansions.
It is then clear that our criterion~\eqref{criterion} is always satisfied, and the isotropization is always realized.
We now see that we can also give general argument for this.

\subsection{General consideration for $\Lambda_0 >0$}

It is known within the framework of classical general relativity that when the BI universe contains
a positive cosmological term $\Lambda_0 > 0$, it asymptotically approaches an isotropically expanding
de Sitter universe. We summarize the argument here for completeness.

The cosmic no hair conjecture asserts that any expanding cosmological spacetimes with positive cosmological
term asymptotically approach, at least locally, the de Sitter universe. This conjecture was proved
in the simple context of the BI cosmologies~\cite{Wald:1983ky} (for further studies on the cosmic no hair conjecture in Bianchi cosmologies, see \cite{Kitada:1991ih,Kitada:1992uh}).
Here, we give a sketch of the proof of the cosmic no hair conjecture, following~\cite{Wald:1983ky}
but restricting our attention to the BI cosmology with $\Lambda_0 > 0$.

We assume that the universe is initially expanding: $H > 0$ at $t = t_0$.
We also impose the dominant energy condition as well as the strong energy condition for our cosmic fluid,~(\ref{eq_EMT}) and~(\ref{eq_EOS}).
From~(\ref{eq:3dotH}),~(\ref{sum:Hi2}), and~(\ref{eq_Ein_BI}),
we obtain the Raychaudhuri equation,
\begin{equation}
\label{eq:Raychaudhuri}
 \dot{H} = - H^2 - \dfrac{\kappa^2}{3 V^2} + \dfrac{\Lambda_0}{3} - \dfrac{4\pi G}{3} (1 + 3 w) \rho \,,
\end{equation}
and the constraint equation,
\begin{equation}
\label{eq:init:constraint}
 H^2 = \dfrac{\Lambda_0}{3} + \dfrac{\kappa^2}{6V^2} + \dfrac{8\pi G}{3} \rho \,,
 % K^2 = 3\Lambda_0 + \dfrac{3}{2} \sigma_{ij} \sigma^{ij} +24 \pi G \rho \geq 0 \,. % 3 \Lambda_0 > 0   \,,
\end{equation}
where $\Lambda_0$ is strictly positive. So we find $H^2 \geq \Lambda_0/3 >0$
and the signature of $H$ never changes so that the universe is expanding for any $t > t_0$. We also find that $V$
is a monotonically increasing function, and the shear squared given by~(\ref{shear2}) is monotonically
decreasing.\footnote{%%%
This can be seen for example, for the Kasner case $V = \ell_0^3 \sinh(3 t/2 \ell_0) \cosh(3 t/2 \ell_0)
\rightarrow (\ell_0 \mathrm{e}^{t/\ell_0})^3$ as $t \rightarrow \infty$ with $\ell_0^2 = 3/\Lambda_0$.
} %%% footnote
As a consequence of the strong energy condition, i.e., $1 + 3 w > 0$, we have from eqs.~(\ref{eq:Raychaudhuri})
and~(\ref{eq:init:constraint})
\begin{equation}
\label{ineq:H}
\dot{H} \leq \dfrac{1}{\ell_0^2} - H^2 \,, %\dot{K} \leq \Lambda_0 - \dfrac{1}{3}K^2 \,.
\end{equation}
with $\ell_0 := \sqrt{3/\Lambda_0}$ being the curvature length of $\Lambda_0$.
This can be integrated to yield\footnote{%%%
More generally, by integrating~(\ref{ineq:H}), we have
$$
H \leq \dfrac{1}{\ell_0} \dfrac{1 + c \, \mathrm{e}^{-2 t/\ell_0}}{1 - c \, \mathrm{e}^{-2 t/\ell_0}} \,,
$$
which is well behaved for all $0< t < \infty$ and yields essentially the same restriction as~(\ref{ineq:H})
at $t \rightarrow \infty$, when $c < 1$.
}
\begin{equation}
 H \leq \dfrac{1}{\ell_0 \tanh (t/\ell_0)} \,.  %{K} \leq \dfrac{ 3/\ell_0}{ \tanh \left(t/\ell_0 \right)}\,,
\end{equation}
Combining again with~(\ref{eq:init:constraint}), we have
\begin{equation}
\dfrac{1}{\ell_0} \leq H \leq \dfrac{1}{\ell_0 \tanh \left(t/\ell_0 \right)} \,.
\label{ineq:Lambda}
\end{equation}
Then, unlike $\Lambda_0 = 0$ case examined in Subsec.~\ref{VA}, $H$ quickly approaches
the nonvanishing constant value $1/\ell_0$ when $t \rightarrow \infty$. This implies that the time dependency
of the spatial metric in the asymptotic future takes the form
\begin{equation}
 g_{ij}(t) = \exp \left( t/\ell_0 \right) g_{ij}(t_0) \,.
\end{equation}
It also follows from~(\ref{eq_Ein_BI_rho}) and $9 H^2 = (\dot{V}/V)^2$ that
\begin{equation}
\label{rho:asympt}
 \rho \leq \dfrac{1}{8 \pi G_0} \left( 3 H^2 - \Lambda_0 \right) \leq \dfrac{1}{8 \pi G_0 \ell_0^2
 \sinh^2\left( t/\ell_0 \right)} \,,
\end{equation}
implying that the cosmic matter also quickly decays in the asymptotic future $t \rightarrow \infty$.
All in all, we conclude that the Binachi-I universe with $\Lambda_0>0$ quickly approaches
an isotropically expanding de Sitter universe in spatially flat cosmological chart, as we see
in the example of the Kasner type solution with positive cosmological term~(\ref{Lambda-Kasner}).
This behavior is called the cosmic no hair.

%%%%%%%%%%%%%%%%%%%%%%%%%%%%%%%%%%%%%%%%%%%%%%%%%%%%%%%%%%%%%%%%%%%%%%
\section{BI Cosmology for $\Lambda_0 > 0$}
\label{BIpL}
%%%%%%%%%%%%%%%%%%%%%%%%%%%%%%%%%%%%%%%%%%%%%%%%%%%%%%%%%%%%%%%%%%%%%%

\subsection{Classical exact solutions}
\label{exact_sol_lambda}

There are four special classical cases that give exact solutions of Eq.~\eqref{solc1} with
$G = G_0$ and $\Lambda = \Lambda_0 > 0$:
\begin{enumerate}
\item
\noindent
$w = 1$:
\begin{equation}
%V_\mathrm{cl}(t) = B_0 \, \cosh\left[ \sqrt{3 \Lambda_0} \, (t - t_0) \right], \qquad
V_\mathrm{cl}(t) = B_0 \, \sinh\left[ \sqrt{3 \Lambda_0} \, (t - t_0) \right],
\end{equation}
where $B_0$ and $t_0$ are integration constants.
The energy density~\eqref{eq_Ein_BI_rho} is given as
\begin{equation}
%\rho_\mathrm{cl}(t) = - \frac{\kappa^2 + 2 B_0^2 \Lambda_0}{16 \pi G_0 B_0^2 \cosh^2\left[ \sqrt{3 \Lambda_0}
% \,(t - t_0) \right]} {\blue = - \frac{2 B_0^2 \Lambda_0 + \kappa^2}{16 \pi G_0 V_\mathrm{cl}^2(t)} }, \qquad
\rho_\mathrm{cl}(t) = \frac{2 B_0^2 \Lambda_0 - \kappa^2}{16 \pi G_0 V_\mathrm{cl}^2(t)}.
\end{equation}
There is another solution involving $\cosh$ instead of $\sinh$ here, but the solution has negative energy density
and is not physically acceptable. This could be understood from the structure of Eq.~\eqref{eq_Ein_BI_rho}:
For the choice of $V \sim \cosh[\sqrt{3 \Lambda_0} \, (t - t_0)]$, its derivative vanishes at $t=t_0$,
giving negative $\rho$ according to Eq.~\eqref{eq_Ein_BI_rho}, and continue to be negative.
Similar solutions exist in the following cases, but they are not exposed here.

\item
$w = 0$:
\begin{eqnarray}
%&& V_\mathrm{cl}(t) = B_0 \, \cosh^2 \left[ \frac12 \sqrt{3 \Lambda_0} \, (t - t_0) \right]
% - \frac{\kappa^2}{2 B_0 \Lambda_0} \exp\left[ \sqrt{3\Lambda_0} (t - t_0) \right],
%\nonumber\\ &&
V_\mathrm{cl}(t) = B_0 \, \sinh^2 \left[ \frac12 \sqrt{3 \Lambda_0} \, (t - t_0) \right]
 - \frac{\kappa^2}{2 B_0 \Lambda_0} \exp\left[ \sqrt{3\Lambda_0} (t - t_0) \right],
\end{eqnarray}
The energy density is given as
\begin{equation}
%\rho_\mathrm{cl}(t) = \frac{B_0^2 \Lambda_0^2}{4 \pi G_0 \left( \kappa^2 \mathrm{e}^{\sqrt{3 \Lambda_0}(t - t_0)}
% - 2 B_0^2 \Lambda_0 \cosh^2 \left[ \frac{1}{2} \sqrt{3 \Lambda_0} \,(t - t_0) \right]\right)}
%{\blue =  - \frac{B_0 \Lambda_0}{8 \pi G_0 V_\mathrm{cl}(t)}, } \qquad
\rho_\mathrm{cl}(t) = \frac{B_0 \Lambda_0}{8 \pi G_0 V_\mathrm{cl}(t)}.
\end{equation}

\item
$w = -1$:
\begin{equation}
%V_{\rm cl}(t) = B_0 \, \sin\left( \sqrt{\frac{3}{2}} \frac{\kappa(t - t_0)}{B_0} \right), \qquad
V_{\rm cl}(t) = B_0 \left( \mathrm{e}^{\frac{t - t_0}{4 B_0}}
- 6 \kappa^2 \mathrm{e}^{-\frac{t - t_0}{4 B_0}} \right),
\end{equation}
The energy density is given as
\begin{equation}
%\rho_\mathrm{cl}(t) = - \frac{\kappa^2 + 2 B_0^2 \Lambda_0}{16 \pi G_0 B_0^2}, \qquad
\rho_\mathrm{cl}(t) = \frac{1 - 48 B_0^2 \Lambda_0}{384 \pi G_0 B_0^2}.
\end{equation}

\item
Any $w$:\\
There is a special solution for any $w$.
\begin{equation}
V(\tau) = B_0 \mathrm{e}^{\sqrt{3 \Lambda_0} \, t} - \frac{\kappa^2}{8 \Lambda_0 B_0}
\mathrm{e}^{- \sqrt{3 \Lambda_0} \, t},
\end{equation}
where $B_0$ is a constant. This has $\rho = 0$.
This is the classical Kasner solution given in Appendix~\ref{kasner_lam}.

\end{enumerate}

\subsection{Power series solutions: classical and {red quantum-improved} for $\Lambda_0 > 0$}
\label{anisotropic_case}
%%%%%%%%%%%%%%%%%%%%%%%%%%%%%%%%%%%%%%%%%%%%%%%%%%%%%%%%%%%%%%%%%%%%%%

We now consider the power series solutions with quantum effects. For this purpose, we need to first
identify the momentum cutoff scale.
In isotropic universe, we have not identified the cutoff scale with the Hubble parameter. The reason is that
it becomes quickly constant~\cite{Chen:2024ebb} for nonzero $\Lambda_0$, and so it is not suitable to describe
the evolution of the universe.

In pursuit of choosing the momentum cutoff scale, we shift into another natural scale, namely the volume
element in BI or the scale factor in isotropic case.
For the case $\Lambda_0 > 0$, the classical solution for isotropic case $\kappa^2 = 0$ can be solved
exactly for any $w$:
\begin{equation}
V_\mathrm{cl}^\mathrm{iso} = D_0 \cosh^{2/(w + 1)}\left[ \frac{w + 1}{2}
 \sqrt{3 \Lambda_0} (t - t_0) \right] \sim D_0  \mathrm{e}^{ \sqrt{3 \Lambda_0} \, t}  ,
\label{ident1}
\end{equation}
where $D_0$ and $t_0$ are integration constants.
Recall again that we are considering only $1 \geq w > -1$.
For BI universe, we observe from the classical exact solutions in Subsec.~\ref{exact_sol_lambda}.
that the leading term of the volume element behaves in the same way as isotropic case for $w \neq -1$,
involving $\kappa^2$ in the integration constant $D_0$. So we may take the same cutoff scale with
the isotropic case discussed in~\cite{Chen:2024ebb} as
\begin{equation}
k = \xi (w + 1) \sqrt{3 \Lambda_0} V^{-1-w}.
\label{ident2}
\end{equation}

To study the power series solutions, Eqs.~\eqref{ident1} and~\eqref{ident2} suggests that we may use
the variable $\tau$ defined by
\begin{equation}
 \tau = \frac1{(w + 1) \sqrt{3 \Lambda_0}} \mathrm{e}^{(w + 1) \sqrt{3 \Lambda_0} \, t}.
\end{equation}
In terms of this variable instead of $t$, Eq.~\eqref{eq_Ein_BI_V} is transformed into
\begin{equation}
\label{eq_V_in_tau}
\frac{\tau^2}{V} \frac{d^2 V}{d\tau^2} + \frac{\tau}{V} \frac{d V}{d\tau}
+ \frac{w - 1}{2} \frac{\tau^2}{V^2} \left( \frac{d V}{d\tau} \right)^2
- \frac{(w - 1) \kappa^2}{4 (w + 1)^2 \Lambda_0} \frac1{V^2} = \frac1{2 (w + 1) \Lambda_0} \Lambda.
\end{equation}
We will now use this equation to obtain both the classical and quantum-improved power series solutions
for the volume element in terms of $\tau$, as done in Subsecs.~\ref{choice} and~\ref{power1}.

To study the power series solution of Eq.~\eqref{eq_V_in_tau}, just as we have done in Subsec.~\ref{power2},
we assume the power expansion solution of $\tau$ in the form
\begin{equation}
V(\tau) = D_0 \tau^s \left( 1 + \frac{D_1}{\tau} + \frac{D_2}{\tau^2} + \cdots \right)
+L_0 \kappa^2 \tau^p \left(1 + \frac{L_1}{\tau} + \frac{L_2}{\tau^2} + \cdots \right) + \mathcal{O}(\kappa^4 \tau^q).,
\label{Tsol_lam}
\end{equation}
where $D_i$'s and $L_i$'s are constants.
The characteristic equation for the leading term $D_0 \tau^{s}$ of power expansion solution is
\begin{equation}
s (s - 1) + s + \frac{w - 1}{2} s^2 - \frac{(w - 1) \kappa^2}{4 (w + 1)^2 \Lambda_0}
 \frac{\tau^{-2s}}{D_0^2} = \frac1{2 (w + 1)}.
\label{chr_lam}
\end{equation}
There are two cases of solutions: (i) $s = 0$ and $D_0^2 = (1 - w) \kappa^2/2 (1 + w) \Lambda_0$
and (ii) $s = 1/(w + 1)$.\footnote{There is another solution $s = -1/(w + 1)$, but it corresponds to contracting
solution which we are not interested in.}
The $\kappa^{2}$ term in the characteristic equation of the leading term hints us to take the series of
$\kappa^2$ term in Eq.~\eqref{Tsol_lam}.  To fix the exponent of $\tau$ in the $\kappa^2$ series,
we write characteristic equation for $V(\tau) = D_0 \tau^s + L_0 \kappa^2 \tau^p$ as
\begin{eqnarray}
&& 2 D_0^2 \Lambda_0 t^{2 s} (w + 1) (s w + s - 1) (s w + s + 1)
\nonumber\\
&+& \kappa^2 \left\{ 1 - w + 4 D_0 \Lambda_0 L_0 (w + 1) t^{p + s}
 \left[ (w + 1) \left( p^2 + p s (w - 1) + s^2 \right) - 1 \right] \right\} + \mathcal{O}(\kappa^4) = 0~.
\end{eqnarray}
From the coefficient of zeroth order in $\kappa$, the exponent of leading order can be fixed as
$s = \pm 1/(w + 1)$ and the exponent of $\tau$ in $\kappa^2$ terms is fixed as $p = -s = \mp 1/(w + 1)$
along with $L_0 = - 1/(8 \Lambda_0 D_0)$. Here we are interested in solutions of expanding universe,
so we will take $s = 1/(w + 1)$ for all $w$ except $w = -1$.
We now analyze each case separately:
\begin{itemize}
\item [(i)]
For $s = 0$, we obtain the classical solution of the volume element in the form
\begin{eqnarray}
V_\mathrm{cl}(\tau) &=& D_0 \left( 1 + \frac{D_1}{\tau^{\sqrt{w+1}}} - \frac{w}{6} \frac{D_1^2}{\tau^{2 \sqrt{w+1}}}
+ \frac{w (w + 1)}{24} \frac{D_1^3}{\tau^{3 \sqrt{w+1}}} - \frac{w (13 w^2 + 30 w + 18)}{1080}
\frac{V_1^4}{\tau^{4 \sqrt{w+1}}} + \cdots \right).
\end{eqnarray}
This solution does not give expanding universe, so it is not what we look for.

\item [(ii)]
The classical solution in this case for all $1 \geq w > -1$ is given as
\begin{eqnarray}
\label{sol_Vcl_Lamba}
V_\mathrm{cl}(\tau) &=& D_0 \tau^\frac1{w + 1} \left\{ 1 + \frac{D_1}{\tau}
- \frac{w - 1}{4} \frac{D_1^2}{\tau^2} - \frac{w (1 - w)}{12} \frac{D_1^3}{\tau^3}
- \frac{w (w - 1) (3 w + 1)}{96} \frac{D_1^4}{\tau^4} + \mathcal{O}\left( \tau^{-5} \right) \right\}
 - \frac{\kappa^2}{8 \Lambda_0 D_0} \tau^{-\frac1{w + 1}}
\nonumber\\ &\times&
 \left\{ 1 + \frac{w (1 - w)}{w + 3} \frac{D_1}{\tau}
- \frac{w (1 - w) (2 w^2 + 3 w + 3)}{4 (w + 2) (w + 3)} \frac{D_1^2}{\tau^2}
+ \mathcal{O}\left( \tau^{-3} \right) \right\}+\mathcal{O}\left(\frac{\kappa^4 D_1}{ D_0^3 \Lambda_0^2}
 \tau^{- (w+4)/(w+1)}\right)~.
\end{eqnarray}
The above solution goes to the Kasner universe for all $w$ for $D_1 = 0$ described in Appendix~\ref{kasner_lam}.
since it leads to the vanishing energy density, namely, $\rho = 0$.

\end{itemize}

The quantum-improved solution could be calculated in the following form:
\begin{eqnarray}
\label{V_qu_ser_lam}
V(\tau) &=& \! D_0 \tau^{\frac{1}{w + 1}} \left\{ 1 \!+\! \frac{D_1}{\tau}
\!-\! \frac{w \!-\! 1}{4} \frac{D_1^2}{\tau^2} \!-\! \frac{ w (1 \!-\! w)}{12} \frac{D_1^3}{\tau^3}
\!-\! \frac{w (w \!-\! 1) (3 w \!+\! 1)}{96} \frac{D_1^4}{\tau^4}
\!-\! \frac{3 (w \!+\! 1) \tilde\mu G_0 \Lambda_0}{4 D_0^{2 w + 2} D_1^{2}} \frac{D_1^2}{\tau^2}
+ \frac{3 w (w \!+\! 1) \tilde\mu G_0 \Lambda_0}{4 D_0^{2 w + 2} D_1^2} \frac{D_1^3}{\tau^3} \right.
\nonumber \\
&-& \left.
\left[ \frac{3 w (w + 1)(3 w + 1) \tilde\mu G_0 \Lambda_0}{16 D_0^{2 w + 2} D_1^2}
- \frac{3 (1 + w)^3 G_0 \Lambda_0 (\tilde\nu + G_0 \Lambda_0 \tilde\mu_1)}{8 D_0^{4 w + 4} D_1^4}
+ \frac{9 w (1 + w)^2 \tilde\mu^2 G_0^2 \Lambda_0^2}{32 D_0^{4 w + 4} D_1^4} \right] \frac{D_1^4}{\tau^4}
+ \mathcal{O}\left( \tau^{-5} \right) \right\}
\nonumber \\
&-& \frac{\kappa^2}{8 \Lambda_0 D_0} \tau^{-\frac1{w + 1}} \left\{ 1 + \frac{w (1-w)}{w + 3} \frac{D_1}{\tau}
- \left[ \frac{w (1 - w) (2 w^2 + 3 w + 3)}{4 (w + 2) (w + 3)}
\!-\! \frac{3 w (w + 1) (2 w + 1)\tilde\mu G_0\Lambda_0}{4 (w + 2) D_0^{2 w + 2} D_1^2} \right] \frac{D_1^2}{\tau^2}
\!+\! \mathcal{O}\left( \tau^{-3} \right) \right\} \nonumber \\
&+& \mathcal{O}\left( \frac{\kappa^4 D_1}{D_0^3 \Lambda_0^2} \tau^{- (w + 4)/(w + 1)} \right).
\end{eqnarray}

In this case too, if the quantum parameters vanish, we recover the classical solution given in~\eqref{sol_Vcl_Lamba}.
Moreover, the $\kappa^2$ term can be combined in the first part when $-2/(w+1)$ is an integer,
i.e., $-2/(w + 1) \in Z \to w = 1, 0,-1/2, -1/3, \cdots$.

\subsection{Numerical analysis of power expanded solutions}
%%%%%%%%%%%%%%%%%%%%%%%%%%%%%%%%%%%%%%%%%%%%%%%%%%%%%%%%%%%%%%%%%%%%%%

\begin{figure}
\includegraphics{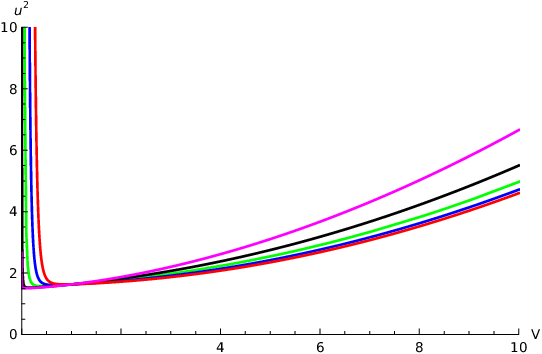}
\caption{The time derivative of volume, $u^2 = \dot V^2$ with quantum correction for $w = 1$ (red), $w = 2/3$ (blue),
$w = 1/3$ (green), $w = 0$ (black) and $w = -1/3$ (magenta) for the chosen parameters.}
\label{fig_expansion}
\end{figure}

To find quantitative difference between the classical and quantum-improved solutions, we will numerically
evaluate the volume element from Eq.~\eqref{sol_eq_x} and estimate
the deviation of the quantum-improved solution from the classical case.

Using the cutoff identification in Eq.~\eqref{ident2} for $\Lambda_0 > 0$, the quantum
correction of the cosmological term is
\begin{eqnarray}
\Lambda(V) = \Lambda_0 - 3 \tilde\mu G_0 \Lambda_0^{2} (1 + w)^2 V^{-2w -2}
+ 9 \left( \tilde\mu_1 G_0^2 \Lambda_0^3 + \tilde\nu G_0 \Lambda_0^2 \right) (1 + w)^4 V^{-4 w - 4} + \cdots,
\end{eqnarray}
where $\tilde\mu = \xi^2 \mu, \tilde\mu_1 = \xi^4 \mu_1$. The quantum-improved equation for $u$ is ($w < 1$)
\begin{eqnarray}
u^2(V) = 3 \Lambda_0 V^2 + c V^{1- w} + \frac32 \kappa^2 + 9 \tilde\mu G_0 \Lambda_0^{2} (1 + w)^2 V^{-2 w}
- 9 \left( \tilde\mu_1 G_0^2 \Lambda_0^3 + \tilde\nu G_0 \Lambda_0^2 \right) (1 + w)^4 V^{-4 w - 2} + \cdots.
\label{lamqu}
\end{eqnarray}
The solution for $V(t)$ can be obtained from Eq.~\eqref{eq_tfromV} applied to \eqref{lamqu}. By solving it numerically,
we can obtain the quantum-improved solutions, while the classical solutions are recovered by setting
$\tilde\mu = \tilde\mu_1 = \tilde\nu = 0$.

To get the idea of the initial condition for $\Lambda_0 > 0$ case, we have plotted the square of the expansion
rate, $u^2(V)$, against the volume element $V$ in Fig.~\ref{fig_expansion}. For the quantum-improved
case, the initial condition is chosen as $V(0) = \mathcal{V}_0$, corresponding to the turning point of $u^2$,
since the universe expands at late times beyond this point.
The behavior of the volume element before the turning point is not reliable, as our analysis is based on
the late-time expansion of the Newton coupling and cosmological term, which is valid only at sufficiently
late times in the study of the BI universe.
On the other hand, in the classical case, $u_\mathrm{cl}^{2}$ exhibits an expanding behavior starting
from the origin. Therefore, we choose the initial condition $V = \mathcal{V}_{0} = 0$ at $t_{0} = 0$.
We then numerically integrate Eq.~\eqref{lamqu} with the parameters
$G_0 = 1, c = 0.1, \kappa = 1, \Lambda_0 = 0.01$ and $\xi = 1$.

\begin{figure}[t]
\begin{minipage}{80mm}
\includegraphics[scale=0.8]{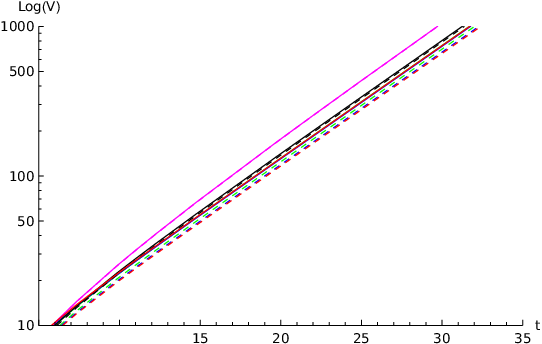}
\caption{The evolution of quantum-improved (solid line) and classical (dashed line) volumes
for $w = 1$ (red), $w = 2/3$ (blue),
$w = 1/3$ (green), $w = 0$ (black) and $w = -1/3$ (magenta) for the chosen parameters.}
\label{f6}
\end{minipage}
\hspace{.5cm}
\begin{minipage}{80mm}
\includegraphics[scale=0.8]{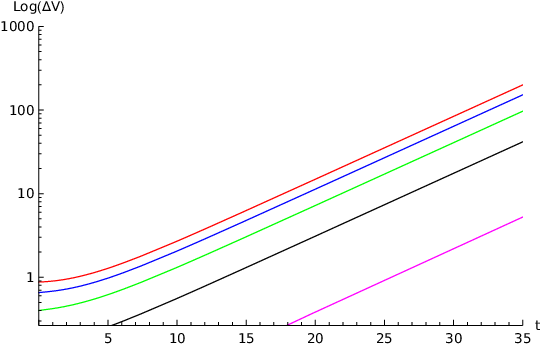}
\caption{The corresponding volume difference $\Delta V = V_\mathrm{qu} - V_\mathrm{cl}$,
in logarithmic scales, for the chosen parameters.}
\label{f7}
\end{minipage}
\end{figure}

We have listed the asymptotic behaviors of the volume element, as well as the differences between
the quantum-improved and classical cases, in Table.~\ref{t2}. It is observed that the differences between
the quantum-improved and classical volume elements follow the same leading-order behavior as the volume element
itself. The results are shown in Fig.~\ref{f6}. This contrasts with the $\Lambda_0 = 0$ case, where the difference captures the next-to-leading
order behavior. To understand the origin of this distinction, we will analyze the cases with $w = 0$ and $w = 1$
in detail.
\begin{enumerate}
\item
For $w = 0$, the expansion rate of volume element~\eqref{lamqu} is given by
\begin{equation}
u^2(V) = 3 \Lambda_0 V^2 + c V + \frac32 \kappa^2 + \eta + \mathcal{O}(V^{-2})
\end{equation}
where we define the quantum correction part as $\eta = 9 \tilde\mu G_0 \Lambda_0^{2}$ for simplification.
Neglecting the higher-order terms $\mathcal{O}(V^{-2})$, the leading-order quantum-improved solution $V(t)$,
obtained from Eq.~\eqref{eq_tfromV}, is approximately
\begin{align}
V_\mathrm{qu} \approx \frac{c + 6 \Lambda_0 \mathcal{V}_0 + \sqrt{6 \Lambda_0} \left[ 2 \eta + 3 \kappa^2
 + 2 \mathcal{V}_0
(c + 3 \Lambda_0 \mathcal{V}_0) \right]^{1/2} }{12 \Lambda_0} \,
\mathrm{e}^{\sqrt{3 \Lambda_0} \, (t - t_0)}
- \frac{c}{6 \Lambda_0} + \mathcal{O}\left( \mathrm{e}^{-\sqrt{3 \Lambda_0}\, (t - t_0)} \right)
\end{align}
where we have imposed the initial condition $V = \mathcal{V}_{0}$ at $t = t_0$. As discussed earlier in
Table~\ref{t2}, the initial volume element $\mathcal{V}_0$ is chosen to be the turning point of $u^{2}$,
and the initial time is fixed as $t_0 = 0$.

On the other hand, in the classical solution, we may impose different initial condition; $V = V_{0}$
at $t = t_0$. Then, the classical solution is obtained as
\begin{align}
V_\mathrm{cl} \approx \frac{c + 6 \Lambda_0 V_{0} + \sqrt{6 \Lambda_0} \left[ 3 \kappa^2
+ 2 V_{0} (c + 3 \Lambda_0 V_{0}) \right]^{1/2}}{12 \Lambda_0} \,
\mathrm{e}^{\sqrt{3 \Lambda_0} \, (t - t_0)} - \frac{c}{6 \Lambda_0} + \mathcal{O}\left(
 \mathrm{e}^{-\sqrt{3 \Lambda_0}\, (t - t_0)} \right)~.
\end{align}
Unlike the $\Lambda_0 = 0$ case, the coefficient of the leading-order term in the solution depends explicitly
on the quantum parameters.
As a result, the difference between the quantum-improved and classical volume elements
does not capture the next to leading order behavior, but instead reflects the difference in the leading-order
terms. This explains why the behavior of $\Delta V$ in Table~\ref{t2} matches the leading-order behavior,
in contrast to Table~\ref{t1}, where $\Delta V$ revealed subleading corrections.

The difference between the quantum-improved and classical volume elements, $\Delta V$, can be expressed
as follows:
\begin{equation}
\Delta V = \left( \frac{\mathcal{V}_{0} - V_{0}}{2} + \frac{\sqrt{2 \mathcal{V}_{0} (c + 3 \Lambda_0 \mathcal{V}_{0})
+ 2 \eta + 3 \kappa ^2} - \sqrt{2 V_{0} (c + 3 \Lambda_0 V_{0})
+ 3 \kappa ^2}}{2 \sqrt{6 \Lambda_0}} \right) \mathrm{e}^{\sqrt{3 \Lambda_0}\, (t - t_0)}
+ \mathcal{O}\left( \mathrm{e}^{-\sqrt{3 \Lambda_0}\, (t-t_0)} \right).
\end{equation}
This difference, $\Delta V$, exhibits the same functional behavior as the leading-order term, but
with a different coefficient. Substituting $\mathcal{V}_{0} = 0.154547$ (from Table~\ref{t2}) and
$V_{0} = 0$ at the initial time $t_0 = 0$, we find $\Delta V = 0.0966722 \, \mathrm{e}^{\sqrt{3 \Lambda_0}\, t}$,
which is consistent with the result shown in Table~\ref{t2}.

It is also important to note that time translation symmetry is preserved in both the classical and quantum-improved cases, as in the $\Lambda_0 = 0$ scenario.

\item
For $w = 1$, we have
\begin{equation}
u^2(V) = 3 \Lambda_0 V^2 + c  + \frac32 \kappa^2 +  4 \eta V^{-2} + \mathcal{O}(V^{-6})~.
\label{diflam2}
\end{equation}
Similarly to the $w = 0$ case, we obtain the quantum-improved solution for the volume element
\begin{align}
V_\mathrm{qu} \approx \frac{2 c + 3\kappa^2}{24 \Lambda_0 \mathcal{C}} \mathrm{e}^{\sqrt{3 \Lambda_0} \, (t - t_0)}
- \mathcal{C} \mathrm{e}^{-\sqrt{3 \Lambda_0}\,(t - t_0)} + \mathcal{O}\left( \mathrm{e}^{-3 \sqrt{3 \Lambda_0} \,
 (t-t_0)} \right).
\end{align}
where $\mathcal{C}$ is the coefficient of the next to leading order term which has the form
\begin{equation}
\mathcal{C} = \frac{\left( 2 c + 3 \kappa ^2 \right) \left(2 c + 3 \kappa^2 + 12 \Lambda_0 \mathcal{V}_0^{2}
- 2 \sqrt{6 \Lambda_0} \sqrt{2 \eta + \mathcal{V}_0^{2} (2 c + 3 \kappa^2 + 6 \Lambda_0
\mathcal{V}_0^{2})} \right)^{1/2}}{2 \sqrt{6 \Lambda_0} \left( 4 c^2 + 12 c \kappa^2 + 9 \kappa^4
- 192 \eta \Lambda_0 \right)^{1/2}}~.
\end{equation}
On the other hand, in the classical solution, imposing different initial condition;
 $V = \mathcal{V}_{0\mathrm{cl}} = 0$ at $t = t_0 = 0$, we
have the classical solution in simple form
\begin{equation}
V_\mathrm{cl} = \frac{\sqrt{2 c + 3 \kappa^2}}{2 \sqrt{6 \Lambda_0}}
\mathrm{e}^{\sqrt{3 \Lambda_0} \,t} - \frac{\sqrt{2 c + 3 \kappa^2}}{2 \sqrt{6 \Lambda_0}}
\mathrm{e}^{-\sqrt{3 \Lambda_0}\, t} + \mathcal{O}\left( \mathrm{e}^{-3\sqrt{3 \Lambda_0}\, t} \right).
\end{equation}
The difference can be calculated from the above solutions by using the initial volume element value
 $\mathcal{V}_{0} = 0.87671$ from Table~\ref{t2} at $t_0 = 0$. This yields
\begin{equation}
\Delta V = 0.464529 \, \mathrm{e}^{\sqrt{3 \Lambda_0} \,t} + \mathcal{O}\left( \mathrm{e}^{-\sqrt{3 \Lambda_0}\, t}
 \right)
\end{equation}
close as Table.\ref{t2}. It is important to note that, for each value of $w$, the leading-order
behavior grows exponentially with a different coefficient. The differences between the cases become apparent
only in the subleading-order terms. The numerical results for $\Delta V$ are shown in Fig.~\ref{f7}.
\end{enumerate}
\begin{table}[tb]
\begin{center}
\begin{tabular}{|c||c|l|l|}
\hline
\quad value of $w$ \quad & \quad initial volume $\mathcal{V}_0$ \quad & \quad asymptotic behavior \quad
& \quad difference of volume $\Delta V$ \quad \\
\hline\hline
$1$ & $0.87671$ & $V \sim 3.98 \times \,  \mathrm{e}^{\sqrt{3 \Lambda_0}\, t}$ & $\Delta V \sim 0.45 \times
 \, \mathrm{e}^{\sqrt{3 \Lambda_0}\, t}$
\\ \hline
$2/3$ & $0.654573$ & $V \sim 4.1 \times \, \mathrm{e}^{\sqrt{3 \Lambda_0}\, t}$ & $\Delta V \sim 0.36 \times \,
  \mathrm{e}^{\sqrt{3 \Lambda_0}\, t}$
\\ \hline
$1/3$ & $0.399989$ & $V \sim 4.23 \times  \, \mathrm{e}^{\sqrt{3 \Lambda_0}\, t}$ & $\Delta V \sim 0.23 \times \,
  \mathrm{e}^{\sqrt{3 \Lambda_0}\, t}$
\\ \hline
$0$ & $0.154547$ & $V \sim 4.45 \times \, \mathrm{e}^{\sqrt{3 \Lambda_0}\, t}$ & $\Delta V \sim 0.10 \times \,
\mathrm{e}^{\sqrt{3 \Lambda_0}\, t}$
\\ \hline
$-1/3$ & $0.013847$ & $V \sim 5.78 \times \,  \mathrm{e}^{ \sqrt{3 \Lambda_0}\, t}$ & $\Delta V \sim .012 \times  \,
 \mathrm{e}^{\sqrt{3 \Lambda_0}\, t}$
\\ \hline
$-2/3$ & $0.0$ & $V \sim 17.78 \times \,  \mathrm{e}^{\sqrt{3 \Lambda_0}\, t}$ & $\Delta V \sim 0.002 \times \,
  \mathrm{e}^{\sqrt{3 \Lambda_0}\, t}$
\\ \hline
\end{tabular}
\end{center}
\caption{Asymptotic behaviors of the total volume and the difference between the quantum-improved
and classical volume element.}
\label{t2}
\end{table}

%%%
\begin{figure}[t]
\includegraphics[scale=1.0]{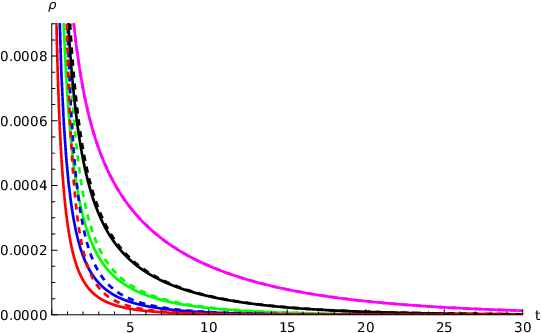}
\caption{The evolution of energy density of quantum-improved (solid) and classical (dashed) solutions
for $w = 1$ (red),
$w = 2/3$ (blue), $w = 1/3$ (green), $w = 0$ (black), $w = -1/3$ (magenta) and $w = -2/3$ (brown)
for the chosen parameters for $\Lambda_0>0$.}
\label{f8}
\end{figure}

We have also calculated the energy density numerically, as shown in Fig. \ref{f8}. We have found that,
in both cases, the energy density diverges at early times and then rapidly decays with time, as expected.

\section{Summary}
\label{summary}

In this work, we have studied the anisotropic BI universe by incorporating quantum corrections
to the Newton coupling and cosmological term within the framework of asymptotically safe gravity.
Our main focus is on the late-time behavior of the universe, which depends on the low-energy flow of the Newton
coupling and the cosmological term in the effective average action under the Einstein-Hilbert truncation.
%We have investigated anisotropy by improving the Einstein equations with the scale dependence
%of the Newton coupling and the cosmological term in place of the the standard Newton and cosmological constants.
In particular, we have employed a modified continuity equation, where only the conservation of the entire
right-hand side of the improved Einstein equations is imposed. The primary aim is to analyze how the anisotropic
universe is influenced by quantum corrections at late times for two distinct branches of FRG trajectories:
one with $\Lambda_0 = 0$ and the other with $\Lambda_0 > 0$.

For studying cosmology in these two branches, we have found that the ``improved'' classical Hubble parameter,
expressed in terms of the quantum-improved scale factor, is a viable cutoff scale choice for the $\Lambda_0 = 0$
branch but not for the $\Lambda_0 > 0$ branch, consistent with our earlier FLRW analysis~\cite{Chen:2024ebb}.
We have derived quantum-improved power series solutions in terms of the cosmic time $t$ for a general equation-of-state
parameter $w$ and observed that these solutions also contain series expansions in the anisotropy parameter
$\kappa^2$ for both branches.

To quantitatively distinguish between the classical and quantum-improved solutions
for general $w$, we have numerically integrated the differential equations and studied the role of initial conditions.
We have shown that, for the $\Lambda_0 = 0$ branch, the coefficients of the leading term and the leading-order
$\kappa^2$ contribution in the power series of $t$ depend only on the parameters in the differential equation and
the equation of state, but not on the initial conditions. The next-to-leading-order terms, however, are
affected by the initial conditions, which govern the differences between the quantum-improved and
classical solutions of the volume element. In contrast, for the $\Lambda_0 > 0$ branch, the difference between
the quantum-improved and classical solutions exhibits the same leading-order behaviors in $\tau$ as
the (classical and quantum-improved) volume elements
themselves. This occurs because, unlike in the $\Lambda_0 = 0$ case, the leading-order coefficients of
the volume for the quantum-improved case explicitly depends on both the quantum parameter and
the initial condition.

We have also examined how the anisotropic Kasner solution is affected by quantum corrections in
Appendix~\ref{kasner_app}.
In the classical Kasner case, the solution is governed by a single integration constant in order for it to
be a solution for any $w$, and it has the vanishing vacuum energy.
In the quantum-improved solution, however, the running cosmological term $\Lambda(t)$ has generated
an effective energy density, which decays with time and vanishes in the asymptotic future.

Moreover we have given a general criterion, valid for any $\Lambda_0$, if the isotropization is achieved
in the late time.  For the $\Lambda_0 = 0$ branch, we have found that for all matter satisfying $-1 < w < 1$,
the universe eventually becomes fully isotropic at late times, since the exponent of the leading behavior
of $V(t)$ is always greater than unity. In the special case $w = 1$, corresponding to stiff matter,
the anisotropic universe may approach the vacuum Kasner solution depending on the behavior of the directional
Hubble parameters.
We have also investigated how quantum effects influence the isotropization of the universe starting from
an initially highly anisotropic state. From the volume element analysis, we have observed that
the quantum-improved volume grows more rapidly than the classical one,
leading to an earlier isotropization of the BI universe due to quantum effects.

For the $\Lambda_0 > 0$ branch, our criterion indicates that the BI universe quickly becomes isotopic.
Indeed we have seen that the cosmic no-hair theorem also holds in the quantum-improved case:
the BI universe quickly approaches an expanding de Sitter universe in a spatially flat cosmological
chart. In particular, the transition from an anisotropic state to a spatially flat de Sitter universe is
found to occur earlier in the quantum-improved case than in the classical scenario.

Although various cosmological observations severely constrain the possibility of anisotropic cosmological models,
the observational viability of the Bianchi Type I universe remains a subject of ongoing discussion
(e.g., \cite{Verma:2023huz}). Furthermore, \cite{Verma:2024lex} discusses its observational viability
with a specific focus on the anisotropy of present-day dark energy.
Our results show that the quantum effects in general accelerate the isotropization except for $w=1$, thereby reducing
the observational constraints and potentially making the realization of such anisotropic dark energy models
more plausible. It would therefore be interesting to consider our quantum-improved BI cosmology along
the lines of these works, and study further observational constraints on AS quantum effects.

\acknowledgments

C.M.C. and N.O. would like to thank Asia Pacific Center for Theoretical Physics for the hospitality where
this work was completed.
R.M. would like to thank the School of Physical Sciences, Indian Association for the Cultivation of Science,
for their hospitality during the completion of this work.
The work of C.M.C. was supported by the National Science and Technology Council of the R.O.C. (Taiwan)
under the grant NSTC 114-2112-M-008-010.
The work of A.I. was supported in part by MEXT KAKENHI Grant-in-Aid for Transformative Research Areas (A)
``Extreme Universe'' No.~JP21H05182, No.~JP21H05186, and by JSPS KAKENHI Grant-in-Aid for Scientific Research
(C) No. 25K07306.
The work of R.M. was supported by the National Science and Technology Council of the R.O.C. (Taiwan)
under the grant NSTC 114-2811-M-008-024.

\begin{appendix}

\section{Kasner solutions}
\label{kasner_app}

\subsection{Classical Kasner solutions for $\Lambda_0 = 0$ and $\rho = 0$}

Here we examine the well-known classical Kasner solution for the $\Lambda_0 = 0$ and $\rho = 0$ case,
which is the vacuum solution of the BI model.
We find the classical exact solution from Eq.~\eqref{eq_Ein_BI_V}:
\begin{eqnarray}
V_\mathrm{cl}^\mathrm{K}(t) = A_0 (t - t_0),
\qquad
a_i(t) = \tilde{a}_{i0} (t - t_0)^{\kappa_i/A_0 + 1/3},
\eea
with
\bea
\sum_{i=0}^3 \kappa_i = 0, \qquad
\kappa^2 = \frac{2}{3}{A_0^2},
\label{consk}
\end{eqnarray}
where $A_0$ and $\tilde{a}_{i0}$ are dimensionfull integration constants.
This is an exact solution for any value of $w$. The price for this is that there is the additional
constraint given in the last relation in~\eqref{consk}.
It seems that this constraint is widely considered to be that on the parameter $\kappa^2$.
However $\kappa^2$ is a parameter in the differential equation~\eqref{eq_Ein_BI_V} we are solving,
and so it is more natural to consider the constant $A_0$ in the solution is constrained rather than $\kappa^2$.
We should regard the Kasner solution as a special one of Eq.~\eqref{eq_Ein_BI_V} with $A_0 := \sqrt{3\kappa^2/2}$
and the differential equation is not constrained.
So there is no constraint on $\kappa_i$ except for the first relation in~\eqref{consk},
and this solution contains only a single integration constant even though it is a solution of the second order
differential equation.

This solution is interesting in that it gives anisotropic solutions for any $w$. This is so
because the solution is given by the power of $t$, consistent with our analysis in Sec.~\ref{sec:IsoBI}.
It is also intriguing that this solution evades the general analysis on the leading power law which indicates
that the power of $t$ should be in general larger than 1 for $-1<w<1$. This is possible because of
the constraint on $V_0$ and the power in $t$.

\subsection{Classical Kasner solutions for $\Lambda_0 > 0$}
\label{kasner_lam}

For the case $\Lambda_0 > 0$,
the governing equation for the classical volume element~\eqref{eq_Ein_BI_V} could be rewritten as
\begin{equation}
\frac{\ddot V}{V} + 12 \pi G (w - 1) \rho = 3 \Lambda_0~.
\label{eq_ein_mod_v}
\end{equation}
As the Kasner solution corresponds to the vanishing energy density $\rho = 0$, we could write a simple solution
for any $w$ from the above equation in the following form
\begin{equation}
V(\tau) = B_0 \mathrm{e}^{\sqrt{3 \Lambda_0} \, t} + B_{1} \mathrm{e}^{- \sqrt{3 \Lambda_0} \, t},
\end{equation}
where $B_0$ and $B_1$  are dimensionless integration constants.
Substituting this solution in the expression of the energy density $\rho$ in Eq.~\eqref{eq_Ein_BI_rho}
and requiring it to vanish, we find that one of the integration constant $V_1$ is constraint as
$B_1 = - \kappa^2/(8 B_0 \Lambda_0)$.
This solution could be written in a compact form choosing
$B_0 = (\kappa/\sqrt{8\Lambda_0}) \, \mathrm{e}^{-\sqrt{3 \Lambda_0} \, t_0}$ as
\begin{eqnarray}
V(t) &=& \frac{\kappa}{\sqrt{2 \Lambda_0}} \sinh\left( \sqrt{3\Lambda_0} \, (t - t_0) \right),
\nonumber\\
a_i(t) &=& a_{i0} \left( \frac{\kappa^2}{2 \Lambda_0} \right)^{\frac{1}{6}}
 \sinh^{1/3}\left( \sqrt{3\Lambda_0} \, (t - t_0) \right) \left[ \tanh\left( \frac{\sqrt{3\Lambda_0}}{2}
(t - t_0) \right) \right]^{\sqrt{\frac{2}{3}} \frac{\kappa_i}{\kappa}}.
\label{Lambda-Kasner}
\end{eqnarray}
Again this is a solution for any $w$ and has only a single integration constant.
Contrary to the Kasner solutions for $\Lambda_0 = 0$, this becomes asymptotically isotropic (de Sitter) universe
after rescaling of the coordinates, in accordance with the general discussions in Sec.~\ref{sec:IsoBI}.
However we should note that these solutions exist only for $\kappa \neq 0$, namely only when there is
an intermediate anisotropy.

\subsection{Quantum-improved Kasner solutions for $\Lambda_0 = 0$}

For the quantum-improved Kasner solution for $\Lambda_0 = 0$, we adopt a simple identification of
the infrared cutoff scale as $k = \xi/t$, which allows us to express the quantum-improved cosmological
term $\Lambda(t)$ using Eq.~\eqref{eq_GLam}. Substituting the general power series expansion~\eqref{gen_power}
into the evolution equation~\eqref{eq_Ein_BI_V}, we obtain the quantum-improved Kasner solution
in the form of a power series as:
\begin{equation}
V^\mathrm{K}(t) = E_0 t \left( 1 - \frac{E_1}{t} - \frac{3 (w + 1) \tilde\nu G_0}{2 (w - 3)} \frac1{t^2}
+ \frac{3 (w + 1) (w - 2) \tilde\nu G_0 E_1}{2 (w - 3) (w - 4)} \frac1{t^3} + \mathcal{O}\left( t^{-4} \right) \right),
\end{equation}
where $E_0$ and $E_1$ are integration constants. As in the classical case, $E_0$ is fixed by
the anisotropy parameter in the same manner. Even though we do not have this solution in a closed form,
we call this quantum-improved Kasner solution because it reduces to the above classical exact solution when
the quantum corrections are turned off.

The energy density for the quantum-improved case is given by
\begin{equation}
\rho(t) = \frac{(w + 1) \tilde\nu}{8 \pi (w - 3)} \frac1{t^4} + \mathcal{O}\left( t^{-6} \right).
\end{equation}
Here, the quantum corrections to $\lambda(t)$ generate an effective energy density that decays with
time and vanishes in the asymptotic future.

\subsection{Quantum-improved Kasner solutions for $\Lambda_0 > 0$}

As discussed in Subsec.~\ref{anisotropic_case} in detail, the classical Kasner solution is obtained for $D_1 = 0$
in Eq.~\eqref{sol_Vcl_Lamba}. The quantum-improved Kasner solution could be found out from
Eq.~\eqref{V_qu_ser_lam} substituting $D_1 = 0$ just like classical case, which leads to the solution as
\begin{eqnarray}
\label{V_qu_kas_lam}
V(\tau) &=& D_0 \tau^{\frac{1}{w + 1}} \left\{ 1
- \frac{3 (w + 1) \tilde\mu G_0 \Lambda_0}{4 D_0^{2 w + 2}} \frac{1}{\tau^2}
+ \left[ \frac{3 (1 + w)^3 G_0 \Lambda_0 (\tilde\nu + G_0 \Lambda_0 \tilde\mu_1)}{8 D_0^{4 w + 4}}
- \frac{9 w(1 + w)^2 \tilde\mu^2 G_0^2 \Lambda_0^2}{32 D_0^{4 w + 4} } \right] \frac{1}{\tau^4}
+ \mathcal{O}\left( \tau^{-5} \right) \right\}
\nn
&-& \frac{\kappa^2}{8 \Lambda_0 D_0} \tau^{-\frac1{w + 1}} \left\{ 1 + \frac{3 w (w + 1) (2 w + 1)
\tilde\mu G_0 \Lambda_0}{4 (w + 2) D_0^{2 w + 2}} \frac{1}{\tau^2}
+ \mathcal{O}\left( \tau^{-3} \right) \right\}.
\end{eqnarray}

\end{appendix}


\begin{thebibliography}{99}

\bibitem{Weinber:1978}
S.~Weinberg,
in: A.~Zichichi (Ed.), ``Understanding the Fundamental Constituents of Matter,'',
(Plenum Press, New York, 1978)

\bibitem{Weinberg}
S.~Weinberg,
``Ultraviolet divergences in quantum theories of gravitation,'',
General Relativity: An Einstein Centenary Survey edited by S.~W.~Hawking and W.~Israel,
(Cambridge University Press, Cambridge, England), Chap. 16 (1979)

%\cite{Reuter:1996cp}
\bibitem{Reuter:1996cp}
M.~Reuter,
``Nonperturbative evolution equation for quantum gravity,''
Phys. Rev. D \textbf{57} (1998), 971-985
%doi:10.1103/PhysRevD.57.971
[arXiv:hep-th/9605030 [hep-th]].

%\cite{Souma:1999at}
\bibitem{Souma:1999at}
W.~Souma,
``Nontrivial ultraviolet fixed point in quantum gravity,''
Prog. Theor. Phys. \textbf{102} (1999), 181-195
%doi:10.1143/PTP.102.181
[arXiv:hep-th/9907027 [hep-th]].

\bibitem{Percacci:2017}
R.~Percacci,
``An Introduction to Covariant Quantum Gravity and Asymptotic Safety,''
100 Years of General Relativity Vol. 3 (World Scientific Publishing, Singapore, 2017)
%ISBN 978-981-320-717-2, 978-981-320-719-6
%doi:10.1142/10369

%\cite{Eichhorn:2018yfc}
\bibitem{Eichhorn:2018yfc}
A.~Eichhorn,
``An asymptotically safe guide to quantum gravity and matter,''
Front. Astron. Space Sci. \textbf{5} (2019), 47
%doi:10.3389/fspas.2018.00047
[arXiv:1810.07615 [hep-th]].

%\cite{Reuter:2019book}
\bibitem{Reuter:2019book}
M.~Reuter and F.~Saueressig,
``Quantum Gravity and the Functional Renormalization Group: The Road towards Asymptotic Safety,''
(Cambridge University Press, 2019)
%ISBN 978-1-107-10732-8, 978-1-108-67074-6
%Cambridge monographs on Mathematical physics.

%\cite{Wetterich:1992yh}
\bibitem{Wetterich:1992yh}
C.~Wetterich,
``Exact evolution equation for the effective potential,''
Phys. Lett. B \textbf{301} (1993), 90-94
%doi:10.1016/0370-2693(93)90726-X
[arXiv:1710.05815 [hep-th]].

%\cite{Morris:1993qb}
\bibitem{Morris:1993qb}
T.~R.~Morris,
``The Exact renormalization group and approximate solutions,''
Int. J. Mod. Phys. A \textbf{9} (1994), 2411-2450
%doi:10.1142/S0217751X94000972
[arXiv:hep-ph/9308265 [hep-ph]].

%\cite{Reuter:2001ag}
\bibitem{Reuter:2001ag}
M.~Reuter and F.~Saueressig,
``Renormalization group flow of quantum gravity in the Einstein-Hilbert truncation,''
Phys. Rev. D \textbf{65} (2002), 065016
%doi:10.1103/PhysRevD.65.065016
[arXiv:hep-th/0110054 [hep-th]].

%\cite{Bonanno:2001xi}
\bibitem{Bonanno:2001xi}
A.~Bonanno and M.~Reuter,
``Cosmology of the Planck era from a renormalization group for quantum gravity,''
Phys. Rev. D \textbf{65} (2002), 043508
%doi:10.1103/PhysRevD.65.043508
[arXiv:hep-th/0106133 [hep-th]].

%\cite{Codello:2008vh}
\bibitem{Codello:2008vh}
A.~Codello, R.~Percacci and C.~Rahmede,
``Investigating the Ultraviolet Properties of Gravity with a Wilsonian Renormalization Group Equation,''
Annals Phys. \textbf{324} (2009), 414-469
%doi:10.1016/j.aop.2008.08.008
[arXiv:0805.2909 [hep-th]].


%\cite{Baldazzi:2021orb}
\bibitem{Baldazzi:2021orb}
A.~Baldazzi and K.~Falls,
``Essential Quantum Einstein Gravity,''
Universe \textbf{7} (2021), no.8, 294
%doi:10.3390/universe7080294
[arXiv:2107.00671 [hep-th]].

%\cite{Ohta:2025xxo}
\bibitem{Ohta:2025xxo}
N.~Ohta and M.~Yamada,
``Essential Renormalization Group Equation for Gravity coupled to a Scalar field,''
[arXiv:2506.03601 [hep-th]].

\bibitem{Chen:2024ebb}
C.-M.~Chen, R.~Mandal and N.~Ohta,
``On the cutoff scale identification of FLRW cosmology in asymptotically safe gravity,''
Class. Quant. Grav. \textbf{42} (2025), no.9, 095011
%doi:10.1088/1361-6382/adcdba
[arXiv:2412.15288 [gr-qc]].


\bibitem{Litim:2001}
D. F. Litim, ``Optimized renormalization group flows",
Phys. Rev. D, {\bf 64}, (2001) 105007.



\bibitem{ReuterWeyer2004}
M.~Reuter and H.~Weyer,
``Quantum gravity at astrophysical distances?,''
JCAP \textbf{12} (2004), 001
[arXiv:hep-th/0410119 [hep-th]].

\bibitem{Kawai:2024rau}
H.~Kawai and N.~Ohta,
``Wave function renormalization in asymptotically safe quantum gravity,''
Phys. Rev. D \textbf{111} (2025), no.4, 046012
[arXiv:2412.08808 [hep-th]].


%\cite{Misner:1967uu}
\bibitem{Misner:1967uu}
C.~W.~Misner,
``The Isotropy of the universe,''
Astrophys. J. \textbf{151} (1968), 431-457
%doi:10.1086/149448

%\cite{Kasner:1921zz}
\bibitem{Kasner:1921zz}
E.~Kasner,
``Geometrical theorems on Einstein's cosmological equations,''
Am. J. Math. \textbf{43} (1921), 217-221
%doi:10.2307/2370192

\bibitem{Heckmann:1959}
O.~Heckmann and E.~Sch\"ucking,
``Newtonsche und Einsteinsche Kosmologie,''
Handbuch der Physik \textbf{53} (1959), 489-519.
%(Springer, Berlin, 1959), p. 489.

%\cite{Khalatnikov:2003ph}
\bibitem{Khalatnikov:2003ph}
I.~M.~Khalatnikov and A.~Y.~Kamenshchik,
``A Generalization of the Heckmann-Sch\"ucking cosmological solution,''
Phys. Lett. B \textbf{553} (2003), 119-125
%doi:10.1016/S0370-2693(02)03237-9
[arXiv:gr-qc/0301022 [gr-qc]].

%\cite{Kamenshchik:2009dt}
\bibitem{Kamenshchik:2009dt}
A.~Y.~Kamenshchik and C.~M.~F.~Mingarelli,
``A Generalized Heckmann-Sch\"ucking cosmological solution in the presence of a negative cosmological constant,''
Phys. Lett. B \textbf{693} (2010), 213-217
%doi:10.1016/j.physletb.2010.08.065
[arXiv:0909.4227 [gr-qc]].

\bibitem{Jacobs:1968}
K.~C.~Jacobs,
``Spatially Homogeneous and Euclidean Cosmological Models with Shear,''
Astrophys. J. \textbf{153} (1968), 661.

\bibitem{Belinskii:1970}
V.~A.~Belinskii, I.~M.~Khalatnikov and E.~M. Lifshitz,
``Oscillatory approach to a singular point in the relativistic cosmology,''
Adv. Phys. \textbf{19} (1970), 525.

%\cite{Hu:1978zd}
\bibitem{Hu:1978zd}
B.~L.~Hu and L.~Parker,
``Anisotropy Damping Through Quantum Effects in the Early Universe,''
Phys. Rev. D \textbf{17} (1978), 933-945
[erratum: Phys. Rev. D \textbf{17} (1978), 3292]
%doi:10.1103/PhysRevD.17.933

%\cite{Cho:1995hz}
\bibitem{Cho:1995hz}
H.-T.~Cho and A.~D.~Speliotopoulos,
``Gravitational waves in Bianchi type I universes. 1: The Classical theory,''
Phys. Rev. D \textbf{52} (1995), 5445-5458
%doi:10.1103/PhysRevD.52.5445
[arXiv:gr-qc/9504046 [gr-qc]].

%\cite{Chen:2000gaa}
\bibitem{Chen:2000gaa}
C.-M.~Chen, T.~Harko and M.~K.~Mak,
``Bianchi type I cosmologies in arbitrary dimensional dilaton gravities,''
Phys. Rev. D \textbf{62} (2000), 124016
%doi:10.1103/PhysRevD.62.124016
[arXiv:hep-th/0004096 [hep-th]].

%\cite{Saha:2001ig}
\bibitem{Saha:2001ig}
B.~Saha,
``Spinor field in Bianchi type I universe: Regular solutions,''
Phys. Rev. D \textbf{64} (2001), 123501
%doi:10.1103/PhysRevD.64.123501
[arXiv:gr-qc/0107013 [gr-qc]].

%\cite{Russell:2013oda}
\bibitem{Russell:2013oda}
E.~Russell, C.~B.~K{\i}l{\i}n{\c{c}} and O.~K.~Pashaev,
``Bianchi I model: an alternative way to model the present-day Universe,''
Mon. Not. Roy. Astron. Soc. \textbf{442} (2014), no.3, 2331-2341
%doi:10.1093/mnras/stu932
[arXiv:1312.3502 [astro-ph.CO]].


\bibitem{Parnovsky:2023}
S. L. Parnovsky,
``The Big Bang could be anisotropic. The case of Bianchi I model,"
	Class. Quant. Grav. {\bf 40} (2023), 135005.

%\cite{Beesham:1994ni}
\bibitem{Beesham:1994ni}
A.~Beesham,
``Bianchi type I cosmological models with variable G and Lambda,''
Gen. Rel. Grav. \textbf{26} (1994), 159-165
%doi:10.1007/BF02105151

%\cite{Kalligas:1995qh}
\bibitem{Kalligas:1995qh}
D.~Kalligas, P.~S.~Wesson and C.~W.~F.~Everitt,
``Bianchi type I cosmological models with variable $G$ and $\Lambda$: A Comment,''
Gen. Rel. Grav. \textbf{27} (1995), 645-650
%doi:10.1007/BF02108066

%\cite{Singh:2007pt}
\bibitem{Singh:2007pt}
J.~P.~Singh, A.~Pradhan and A.~K.~Singh,
``Bianchi type-I cosmological models with variable $G$ and $\Lambda$-term in general relativity,''
Astrophys. Space Sci. \textbf{314} (2008), 83-88
%doi:10.1007/s10509-008-9742-6
[arXiv:0705.0459 [gr-qc]].

%\cite{Pradhan:2013jg}
\bibitem{Pradhan:2013jg}
A.~Pradhan, R.~Jaiswal and R.~K.~Khare,
``Bianchi type-I cosmological models with time dependent $q$ and $\Lambda$-term in general relativity,''
Astrophys. Space Sci. \textbf{343} (2013), 489-497
%doi:10.1007/s10509-012-1239-7

%\cite{Mandal:2019xlg}
\bibitem{Mandal:2019xlg}
R.~Mandal, S.~Gangopadhyay and A.~Lahiri,
``Cosmology of Bianchi type-I metric using renormalization group approach for quantum gravity,''
Class. Quant. Grav. \textbf{37} (2020), no.6, 065012
%doi:10.1088/1361-6382/ab7287
[arXiv:1906.08674 [gr-qc]].


\bibitem{Platania_2020}
A. Platania,
``From renormalization group flows to cosmology",
Front. in Phys. {\bf{8}} (2020), 188 [arXiv:2003.13656 [gr-qc]].


%\cite{Reuter:2003ca}
\bibitem{Reuter:2003ca}
M.~Reuter and H.~Weyer,
``Renormalization group improved gravitational actions: A Brans-Dicke approach,''
Phys. Rev. D \textbf{69} (2004), 104022
%doi:10.1103/PhysRevD.69.104022
[arXiv:hep-th/0311196 [hep-th]].

%\cite{Bonanno:2001hi}
\bibitem{Bonanno:2001hi}
A.~Bonanno and M.~Reuter,
``Cosmology with selfadjusting vacuum energy density from a renormalization group fixed point,''
Phys. Lett. B \textbf{527} (2002), 9-17
%doi:10.1016/S0370-2693(01)01522-2
[arXiv:astro-ph/0106468 [astro-ph]].

%\cite{Bonanno:2007wg}
\bibitem{Bonanno:2007wg}
A.~Bonanno and M.~Reuter,
``Entropy signature of the running cosmological constant,''
JCAP \textbf{08} (2007), 024
%doi:10.1088/1475-7516/2007/08/024
[arXiv:0706.0174 [hep-th]].

%\cite{Hindmarsh:2011hx}
\bibitem{Hindmarsh:2011hx}
M.~Hindmarsh, D.~Litim and C.~Rahmede,
``Asymptotically Safe Cosmology,''
JCAP \textbf{07} (2011), 019
%doi:10.1088/1475-7516/2011/07/019
[arXiv:1101.5401 [gr-qc]].

%\cite{Reuter:2004nv}
\bibitem{Reuter:2004nv}
M.~Reuter and H.~Weyer,
``Running Newton constant, improved gravitational actions, and galaxy rotation curves,''
Phys. Rev. D \textbf{70} (2004), 124028
%doi:10.1103/PhysRevD.70.124028
[arXiv:hep-th/0410117 [hep-th]].

%\cite{Mandal:2020umo}
\bibitem{Mandal:2020umo}
R.~Mandal, S.~Gangopadhyay and A.~Lahiri,
``Cosmology with modified continuity equation in asymptotically safe gravity,''
Eur. Phys. J. Plus \textbf{137} (2022), 10
[arXiv:2010.09716 [gr-qc]].


%\cite{Barrow:1987ia}
\bibitem{Barrow:1987ia}
J.~D.~Barrow,
``Cosmic No Hair Theorems and Inflation,''
Phys. Lett. B \textbf{187}, 12-16 (1987).
%doi:10.1016/0370-2693(87)90063-3
%258 citations counted in INSPIRE as of 24 Oct 2025

%\cite{Kitada:1991ih}
\bibitem{Kitada:1991ih}
Y.~Kitada and K.~Maeda,
``Cosmic no hair theorem in power law inflation,''
Phys. Rev. D \textbf{45}, 1416-1419 (1992).
%doi:10.1103/PhysRevD.45.1416
%125 citations counted in INSPIRE as of 24 Oct 2025

%\cite{Kitada:1992uh}
\bibitem{Kitada:1992uh}
Y.~Kitada and K.~Maeda,
``Cosmic no hair theorem in homogeneous space-times. 1. Bianchi models,''
Class. Quant. Grav. \textbf{10}, 703-734 (1993).
%doi:10.1088/0264-9381/10/4/008
%118 citations counted in INSPIRE as of 24 Oct 2025
%%% color

\bibitem{Brizuela:2024}
D. Brizuela and S. F. Uria,
``Analytic solutions for the Bianchi I universe coupled to several barotropic perfect fluids,''
Phys. Rev. D \textbf{110} (2024), 084059.

\bibitem{Reuter:2005jcap}
M. Reuter and F. Saueressig,
``From big bang to asymptotic de Sitter: complete cosmologies in a quantum gravity framework'',
JCAP {\bf 09} (2005), 012 [arXiv:hep-th/0507167].

\bibitem{BF_2017}
A. Bonanno and F. Saueressig,
``Asymptotically safe cosmology - a status report",
Comptes Rendus Physique {\bf{18}}, (2017), 254 [arXiv:1702.04137 [hep-th]].

\bibitem{Moti:2019}
R. Moti and A. Shojai,
``On the cutoff identiﬁcation and the quantum improvement in
asymptotically safe gravity", Phys. Lett. B {\bf 793}  (2019), 313.


%\cite{Bronnikov:2004nu}
\bibitem{Bronnikov:2004nu}
K.~A.~Bronnikov, E.~N.~Chudaeva and G.~N.~Shikin,
``Magneto-dilatonic Bianchi I cosmology: Isotropization and singularity problems,''
Class. Quant. Grav. \textbf{21} (2004), 3389-3403
%doi:10.1088/0264-9381/21/14/005
[arXiv:gr-qc/0401125 [gr-qc]].

%\cite{Saha:2006iu}
\bibitem{Saha:2006iu}
B.~Saha,
``Nonlinear spinor field in Bianchi type-I cosmology: Inflation, isotropization, and late time acceleration,''
Phys. Rev. D \textbf{74} (2006), 124030
%doi:10.1103/PhysRevD.74.124030

%\cite{Wald:1983ky}
\bibitem{Wald:1983ky}
R.~M.~Wald,
``Asymptotic behavior of homogeneous cosmological models in the presence of a positive cosmological constant,''
Phys. Rev. D \textbf{28} (1983), 2118-2120
%doi:10.1103/PhysRevD.28.2118

%\cite{Verma:2023huz}
\bibitem{Verma:2023huz}
A.~Verma, S.~K.~Patel, P.~K.~Aluri, S.~Panda and D.~F.~Mota,
``Constraints on Bianchi-I type universe with SH0ES anchored Pantheon+ SNIa data,''
JCAP \textbf{06} (2024), 071
%doi:10.1088/1475-7516/2024/06/071
[arXiv:2310.07661 [astro-ph.CO]].
%3 citations counted in INSPIRE as of 09 Oct 2025

%\cite{Verma:2024lex}
\bibitem{Verma:2024lex}
A.~Verma, P.~K.~Aluri and D.~F.~Mota,
``Anisotropic universe with anisotropic dark energy,''
Phys. Rev. D \textbf{111} (2025), no.8, 083508
%doi:10.1103/PhysRevD.111.083508
[arXiv:2408.08740 [astro-ph.CO]].
%4 citations counted in INSPIRE as of 09 Oct 2025




\end{thebibliography}
\end{document}